\documentclass[structabstract]{aa}  
\usepackage{graphicx}
\usepackage{lscape}
\usepackage{amsmath}
\usepackage{txfonts}
\usepackage{url}
\usepackage{natbib}
\usepackage{xcolor}
\newcommand{\teff}{\ensuremath{\mathit{T}_{\rm eff}}}                % effective temperature

\begin{document}
   \title{The impact of mass-loss on the evolution and pre-supernova properties of red supergiants}
%   \title{Red supergiants as constraints and challenges for stellar evolution: impact of mass-loss, rotation and metallicity}
%\title{Red supergiants and post red supergiants stars: impact of mass-loss and rotation at solar metallicity}
 
   \author{G. Meynet\inst{1},
   V. Chomienne\inst{1},   
   S. Ekstr\"om\inst{1},
   C. Georgy\inst{2},
   A. Granada\inst{1},
   J. Groh\inst{1},  
   A. Maeder\inst{1},    
    P. Eggenberger\inst{1},     
    E. Levesque\inst{3}   
    \and
   P. Massey\inst{4}      
}

   \authorrunning{Meynet et al.}

 \institute{Geneva Observatory, University of Geneva, Maillettes 51, CH-1290 Sauverny, Switzerland
                    \and Astrophysics, Lennard-Jones Laboratories, EPSAM, Keele University, Staffordshire ST5 5BG, UK
           \and CASA, Department of Astrophysical and Planetary Sciences, University of Colorado 389-UCB, Boulder, CO 80309, USA                    
           \and Lowell Observatory, 1400 W Mars Hill Road, Flagstaff, AZ 86001, USA}

   \date{Received ; accepted }

% \abstract{}{}{}{}{} 
% 5 {} token are mandatory
 \abstract
  % context heading (optional) leave it empty if necessary  
   {The post main-sequence evolution of massive stars is very sensitive to many parameters of the stellar models. Key parameters are the mixing processes, the metallicity, the mass-loss rate and the effect of a close companion.} 
   {We study how the red supergiant lifetimes, the tracks in the Hertzsprung-Russel diagram (HRD), the positions in this diagram of the pre-supernova progenitor as well as the structure
   of the stars at that time change for various mass-loss rates during the red supergiant phase (RSG), and for two different initial rotation velocities.}
   {Stellar models are computed with the Geneva code for initial masses between 9 and 25 M$_\odot$ at solar metallicity (Z=0.014) with 10 times and 25 times the standard mass-loss rates during the red supergiant phase, with and without rotation.}
   {
% Enhanced mass-loss during the red supergiant phase of the 9M$_\odot$ produces an increase of the RSG lifetime, while in general for more massive stars, it produce a decrease, which can be up to an order of magnitude. This difference comes from the fact that below the mass limit, the increased mass-loss suppresses the blue loop, while above this mass limit, increased mass-loss produces a blueward evolution. Interestingly, the most massive stars cannot lose more than a maximum amount of mass during the RSG phase. This comes from the fact that the higher the RSG mass-loss, the shorter the phase is, keeping the total mass lost below a given level. The luminosity function along the red supergiant branch is sensitive to the mass-loss rates during that phase. Enhanced mass-loss tend to produce lower luminous and bluer progenitors for core collapse progenitors, a point already emphasized in previous works. 
The surface abundances of RSGs are much more sensitive to rotation than to the mass-loss rates during that phase.
A change of the RSG mass-loss rate has a strong impact on the RSG lifetimes and therefore on the luminosity function of RSGs.  An observed RSG is associated to a larger initial mass model, when enhanced RSG mass-loss rate models are used to deduce that mass.
 %We show that the following properties of the RSG, positions in the HRD, radii, surface abundances and surface velocities are mostly insensitive to the RSG mass-loss rate used.
%A star leaves the RSG region when the envelope mass is reduced to a certain value for a given core mass, this implies that whatever the mass loss rate used, the maximum quantity of mass that can be lost during the RSG phase remains the same.
%For a star with an initial mass between 15 and 25 M$_\odot$, the maximum mass that
%can be lost during the RSG phase is at most 40-60\% of the initial mass.
%Arguments based on the positions of the red supergiant progenitors of type IIP supernovae 
%The transition between red supergiants being the end point of the evolution and red supergiants being in a transitory phase could be around 17 M$_\odot$ at solar metallicity. This would explain why Smartt (2009). 
%and on the the RSG populations in clusters indicate that
%the standard mass-loss rate are likely appropriate for describing the bulk of the RSG populations. On the other hand, the existence of yellow or blue progenitors with initial masses between 15 and 25 M$_\odot$ favors RSG enhanced mass-loss rates. 
At solar metallicity, the enhanced mass-loss rate models do produce significant changes on the populations of blue, yellow and red supergiants. When extended blue loops or blue ward excursions are produced by enhanced mass-loss, the models predict that a majority of blue (yellow) supergiants are post RSG objects. These post RSG stars are predicted to show much smaller surface rotational velocities than similar blue supergiants on their first crossing of the HR gap.
Enhanced mass-loss rates during the red supergiant phase has little impact on the Wolf-Rayet (WR) populations. 
The position in the HRD of  the end point of the evolution depends on the mass of the hydrogen envelope. More precisely, 
whenever, at the pre-supernova stage, the H-rich envelope contains more than about 5\% of the initial mass, the star is a red supergiant, and whenever the H-rich envelope contains less than 1\% of the total mass
the star is a blue supergiant. For intermediate situations, intermediate colors/effective temperatures are obtained. Yellow progenitors for core collapse supernovae can be explained
by the enhanced mass-loss rate models, while the red progenitors are better fitted by the standard mass-loss rate models.
   }
{}
%   {Changing the mass-loss rates during the red supergiant phase has many consequences for the evolution of massive stars and will have a consequence also on the
%   expected luminosity function of red suoergiant stars. Higher the mass-loss, steeper will be the decline of the luminosity function of the red supergiants in a constant star formation %system.}
 

   \maketitle
%==================================================================
%__________________________________________________________________

\section{Introduction}

Red supergiants (RSG) represent a long-lasting stage during the
core He-burning stage of all massive stars with masses between about 9 and 25 M$_\odot$. 
Therefore, a large number of post-Main Sequence massive stars are expected to be in this evolutionary stage.
This kind of stars represent the end point of the evolution of about half of the massive stars. When such a star core-collapses, it produces a type II-P or type II-L supernova event.
%\footnote{Some stars exploding as red supergiants may have so little amount of H in their envelope that they may explode as type II-L supernovae.
%Interactions of the supernova ejecta with the dense surrounding red supergiant wind may also produce type IIn supernova.}. 
Interactions of the supernova ejecta with the dense red supergiant wind may in some circumstances produce type II-n supernovae \citep[see e.g.][]{smith2011}.
These stars are the progenitors of neutron stars and maybe also of some black holes. Thanks to their high luminosities, 
they can be observed far away in the Universe and used to probe, for instance, the metallicity of distant galaxies \citep{Davies2010}. 
For all these reasons, red supergiants represent key objects to understand.

While models can reproduce satisfactorily some observed properties of red supergiants as for instance their positions in the HR diagrams \citep{Levesque2005, Levesque2006, Massey2009}, these stars 
also pose some interesting questions. For instance, 
%Recently, \citet{Davies2013} has suggested that the effective temperatures of RSG are underestimated. 
%\citet{Dessart2013} proposes on the basis of the type II-SN light curves that the radii of RSG are smaller than usually thought.
%Many works address the question of what is the final evolution of those stars, both
%from a theoretical \citep{Salasnich1999, Vanbeveren2007, Yoon2010, Georgy2012} and an observational point of view \citep{Humphreys1991, Humphreys2013}. 
are all red supergiants exploding in a type II core collapse event?
Or do some of them represent a transitory stage before the star evolves back to bluer regions of the HR diagram \citep{Yoon2010,Georgy2012WR,Georgy2012,gme13,Groh2013}? 
Can such further evolution explain  the high number of blue supergiants observed in solar metallicity clusters with mass at the turn off around 9-15 M$_\odot$ \citep{Meylan1983, Eggenberger2002}, 
the low-luminosity WC stars \citep{Georgy2012WR}, low-luminosity luminous blue variables (LBV) as SN progenitors \citep{gme13} and/or the 
yellow supergiants progenitors of core collapse events \citep{Georgy2012}, as for instance 2011dh \citep{VAN11dh2013}?

Actually all these questions are somewhat intertwined and should be addressed simultaneously. This is what we want to do in the present paper, focusing on solar metallicity models.
A key point for answering these questions is a good knowledge of the mass-loss rates during the RSG phase.
Unfortunately this quantity is not well constrained observationally (see the discussion in the next section). 
%Typically, at solar metallicity, single star models from \citet{Ekstrom2012} show that a 20 M$_\odot$  star loses about 40\%
%of its initial mass. The RSG wind mass-losses have a strong impact on the circumstellar environment of these stars, their possible reddening 
%(the cooling wind is likely the site of dust formation), and for determining the end of their evolution (evolution back to the blue or not), 
%the structure and the properties of the star just before the core collapses, therefore the type of the SN event.
%%%%%+++A METTRE PLUS LOIN
%On the other hand, these mass-losses have likely no very important consequences for the chemical enrichment of the interstellar medium.
%The wind ejected material is mainly enriched in nitrogen, 
%however in the solar neighborhood most of the nitrogen is produced by the more numerous intermediate mass stars. Moreover, the loss of the envelope does
%not much affect the evolution of the central regions that would follow the same evolution for many different RSG mass-loss rates.
%Therefore, the composition of the SN ejecta in oxygen for instance are not much affected by different mass-loss histories during the RSG phase. 
The difficulty to deduce the mass-loss rates during the RSG phase comes from the complexity of the envelopes of these stars
and thus from the difficulty to obtain reliable spectroscopic diagnostics. 
Moreover, it is likely that these stars do not lose mass uniformly as a function of time but undergo strong and short outbursts.  
%{\bf DO WE NEED THIS NEXT SENTENCE? IT MAY BE CONFUSING FOR SOME PEOPLE In that respect, 
%red supergiants may be some kinds of siblings, in the red part of the HR diagram, of 
%the luminous blue variables.}
One spectacular example is VY CMa. 
It is a luminous M supergiant with a luminosity equal to 2-3 $\times$ 10$^5$ L$_\odot$  
for the parallax distance of 1.14$\pm$0.09 kpc  \citep[see][]{Choi08}.
The star has a dusty circumstellar envelope \citep{Humphreys07} which produces a reflection nebula at optical wavelengths. \citet{Smith01} estimate the mass of the nebula surrounding VY CMa as 0.2-0.4 M$_\odot$.
%Using Herschel, \citet{Matsuura14} analyzed the far IR and submillimiter molecular emission line spectrum of the M-supergiant VY CMa.
%They fit the line fluxes of $^{12}$CO, $^{13}$CO, H$_2$O and SiO with a mass-loss rate of 1.85 $\times$ 10$^{-4}$ M$_\odot$ yr$^{-1}$.
%This is in agreement with previous estimates by \citet{Danchi94} and
%\citet{DeBeck10}. 
\citet{Decin06} deduced that VY CMa had undergone a phase of high mass-loss (about 3.2 $\times$ 10$^{-4}$ M$_\odot$ yr$^{-1}$) some 1000 years ago (see Fig.~\ref{fig1VL}).

If this outbursting mode of losing mass was common among red supergiants, then this would make their modeling quite difficult \footnote{Note that strong mass losses may also be triggered by mass transfer in a close binary system during the RSG phase. Our enhanced mass-loss rate models during the RSG phase may mimic such
a situation.}.
Even with reliable spectroscopic diagnostics of the current mass-loss rate, our vision might be biased towards mass-loss rate values representative of the long periods when the star is in a weak wind regime.
We may miss the much more seldom and short mass-loss episodes during which most of the mass might be lost. Moreover,
we have at present no firm theory for making predictions on the frequency and the durations of such outbursts, although some authors have proposed that they may be triggered by
pulsations \citep{Yoon2010}. So, at the moment, the most promising way to make progress is to compute models with various mass-loss rates during the RSG stage and to see whether
some range of mass-losses seem to be preferred over others to predict some peculiar outcomes.
This is the aim of the present work, where we build on initial efforts from \citet{Georgy2012} that focused on the pre-SN phase.

% The present study focuses on the impact of high mass-loss rate regimes that seems to be required to 
%reproduce the core collapse events arising from yellow or blue progenitors (see section NN below).
%In this work, we consider the average red supergiant mass-loss rate as a free parameter of the model.
%Our aim will be to investigate the consequences of various mass-loss rates on the properties of red and post-red supergiants.
%and to investigate the consequences of various mass-loss rate values on various properties of red supergiants and post-red supergiants.  
%One of the aim will be to identify the possible ranges for the mass-losses during the RSG phase. Ideally, 
%we would like to know the percentages of RSG characterized by a given mass
%loss rate and of course the physical reasons for this peculiar occurrence of the stellar winds. As we shall see this last goal does appear
%beyond what is possible today but we shall indicate how it may be possible to make some steps further in that direction.
In Sect. 2 we present the physical ingredients of our stellar models. 
Section 3 discusses the impact of different RSG mass-loss rates on the evolutionary tracks and lifetimes. The impacts on the properties of the RSG and of the post-RSG are investigated in Sects. 4 and 5 respectively. 
Implications for the populations of Wolf-Rayet, blue, yellow and red supergiants are discussed in Sect. 6. 
Conclusions and perspectives are presented in Sect. 7.

 \section{The stellar models \label{SecPhymod}}
   
Except for the mass-loss rates in the non-standard cases (see below), the models are computed with exactly the same physical ingredients as the models computed by \citet{Ekstrom2012}, so the interested reader can refer to that paper for a detailed account. Let us just recall here that the models are computed with the Schwarzschild criteria for convection with a core overshooting. The core extension due to overshooting is taken equal to 10\% the pressure scale height estimated at the Schwarzschild core boundary. Non adiabatic convection is considered in the outer convective zone with a mixing length scale equals to 1.6 times the local pressure scale height. 
In rotating models, the shear turbulence coefficient is taken from \citet{Maeder1997}, while the horizontal turbulence and the effective diffusion coefficients are those from \citet{Zahn1992}.

The mass-loss prescription for the hot part of the evolutionary tracks is that of  \citet{dejager1988} for the initial masses 9 and 15 M$_\odot$ and for Log $(T_{\rm eff}/K) > 3.7$. For Log $(T_{\rm eff}/K) < 3.7$, we use a fit on the data by \citet{Syl1998}
and \citet{vanLoon2005} as suggested by \citet{Crowther2001}. Above 15 M$_\odot$, the prescription given by \citet{Vink2001} is used on the MS phase as long as Log $(T_{\rm eff}/K) > 3.9$, the recipe from \citet{dejager1988} is used
for the non red supergiant phase. For Log $(T_{\rm eff}/K) < 3.7$, the prescription is the same as for lower initial mass stars.
The effects of rotation on the mass-loss rates are accounted for as in \citet{mm6}. 
Note that these effects are quite negligible for the rotation rates considered in this work.

As explained in \citet{Ekstrom2012}, for massive stars ($>15\,\text{M}_{\sun}$) in the red supergiant phase, some points in the most external layers of the stellar envelope might exceed the Eddington luminosity of the star: $L_\text{Edd} = 4\pi cGM/\kappa$ (with $\kappa$ the opacity). This is due to the opacity peak produced by the variation of the ionization level of hydrogen beneath the surface of the star. 
We account for this phenomenon by increasing the mass-loss rate of the star (computed according to the prescription described above) by a factor of $3$. Once the supra-Eddington layers disappear, later during the evolution, we come back to the usual mass-loss rate. Adopting this factor 3 in case of supra-Eddington luminosity layers and the mass loss recipes indicated above produces the standard time-averaged mass loss rates shown by the heavy continuous lines in Fig.~\ref{fig1VL}. One sees that it well goes along the average mass loss rate determinations for RSG's by various authors.

For the enhanced mass-loss rate models, we just multiply by a factor 10 or 25 the mass-loss rates as given by the prescriptions indicated above during the whole period when the star is a RSG.
We consider the star is a RSG when its effective temperature (\teff), as estimated by the Geneva code, is $\log (\teff/K) < 3.7$. We chose this limit because
for every stellar models considered here, it encompasses the evolutionary phase during which the tracks become vertical in the HR diagram.  Although this is a slightly too high \teff\ for RSGs, using a smaller limiting value such as $\log (\teff/K)< 3.6$ would not have changed our results. This is because the part
of the evolution comprised between $\log (\teff/K) = 3.6$ and 3.7 is very short compared to the time spent with $\log (\teff/K) < 3.6$.
Note that when these enhanced mass loss rates are used, we do not account for the effect of the supra-Eddington layers as described in  \citet{Ekstrom2012}. This means that the enhancement of the mass-loss rates with respect to those used in \citet{Ekstrom2012} are actually a little less than the factor 10 and 25. The reader can look at Fig.~\ref{fig1VL} to get an idea of the enhancement factor with respect to the mass loss rate used in \citet{Ekstrom2012}.

The enhancement factors of 10 and 25 are chosen somewhat arbitrarily. The only guideline we considered were to not overcome the maximum mass loss rates estimated by  \citet{vanLoon2005} (see  the empty circles in Fig.~\ref{fig1VL}). 
%Let us remind here that the spectroscopically determined mass-loss rates are obtained at a given time and does not say anything about the duration of the period during which such mass loss rates are valid.
%one does not know whether the estimated mass loss rates are active during the whole RSG phase or not or whether they are varying with time. 
We could have explored also the cases of mass loss rates lower than
those indicated by the continuous line in Fig.~\ref{fig1VL}. In that case, the differences with respect to the standard tracks presented here would be mainly an increase of the maximum initial mass of stars that end
their lifetimes as RSGs. 
%Although, we cannot at present discard the possibility that such tracks would well correspond to some cases, we think that 
The cases with enhanced mass loss rates are the most interesting to study
since they can propose a solution for those stars, in the mass range between 9 and 25 M$_\odot$, ending their lifetimes as yellow, blue or even Wolf-Rayet stars, moreover such models somewhat mimics what would be the evolution in case
of strong mass losses, during the RSG phase, triggered by a mass transfer in a close binary system. That is why we concentrate on those models in the present work.

\subsection{Comparisons with spectroscopically determined mass-loss rates}

\begin{figure}[h]
\centering
\includegraphics[width=.47\textwidth]{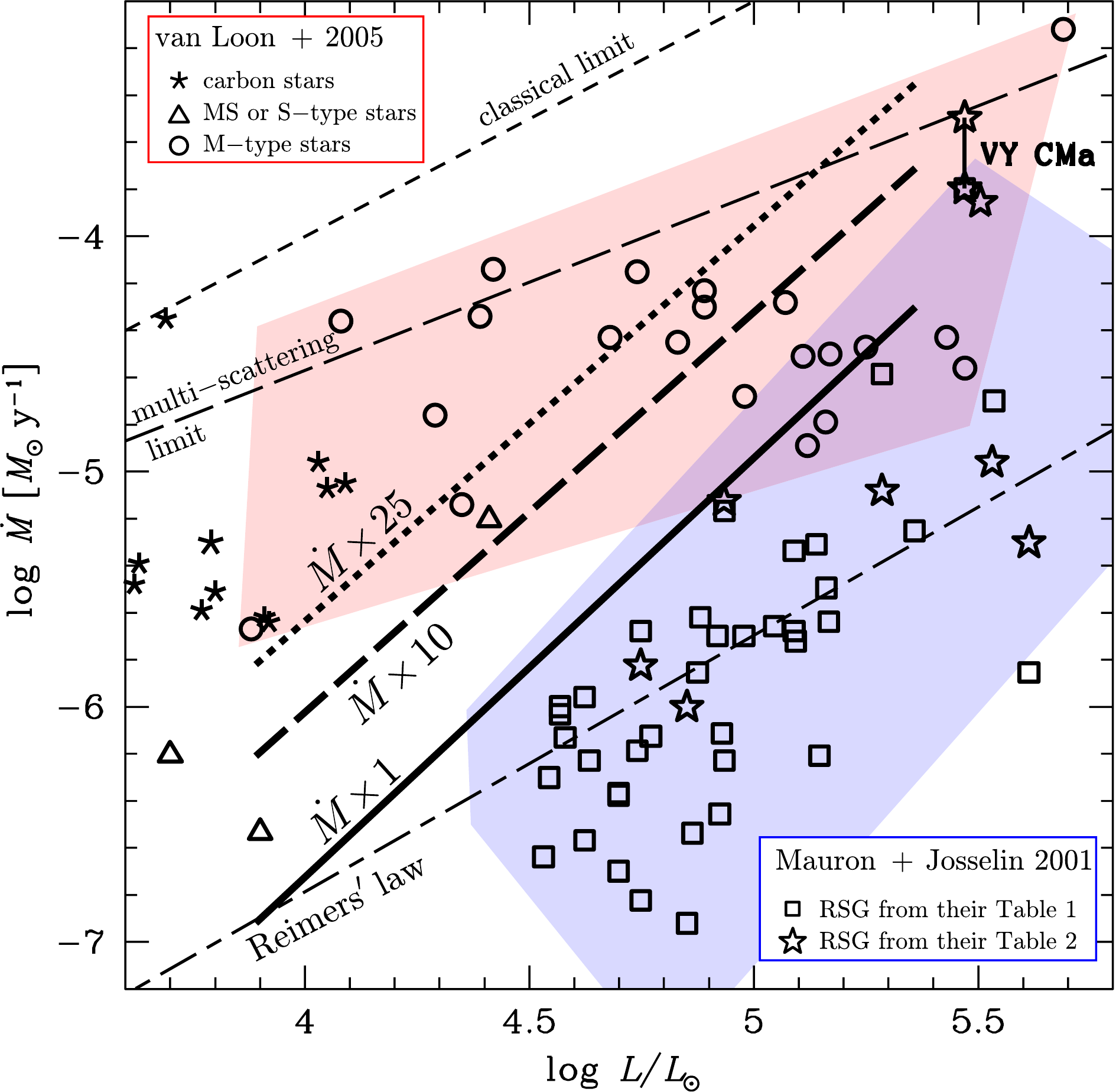}
   \caption{Comparison between the mass-loss rates deduced from spectroscopy by different authors and the average values used in our models during the red supergiant phase. The data by  \citet{vanLoon2005} cover the region shaded in red. The data by \citet{Mauron2011} are distributed in the region shaded in blue.
The two stars linked by a vertical line show the mass-loss rates  for VY CMa; the lower value is taken from the table 1 of \citet{Mauron2011}, the upper one is from \citet{Matsuura14}.The heavy continuous, long-dashed and dotted lines are our averaged red supergiant mass-loss rates equal to respectively 1, 10 and 25 times the standard mass-loss rate (see text). The upper thin
long and short-dashed slanted lines mark the classical and
multiple-scattering limits to the mass-loss rate \citep{vanLoon1999}.  The lower thin short-long-dashed line shows the Reimers law for an average temperature T$_{\rm eff}$= 3750 K  \citep[see][]{Mauron2011}.
   }
      \label{fig1VL}
\end{figure}

 \begin{figure}[h]
\centering
\includegraphics[width=.47\textwidth]{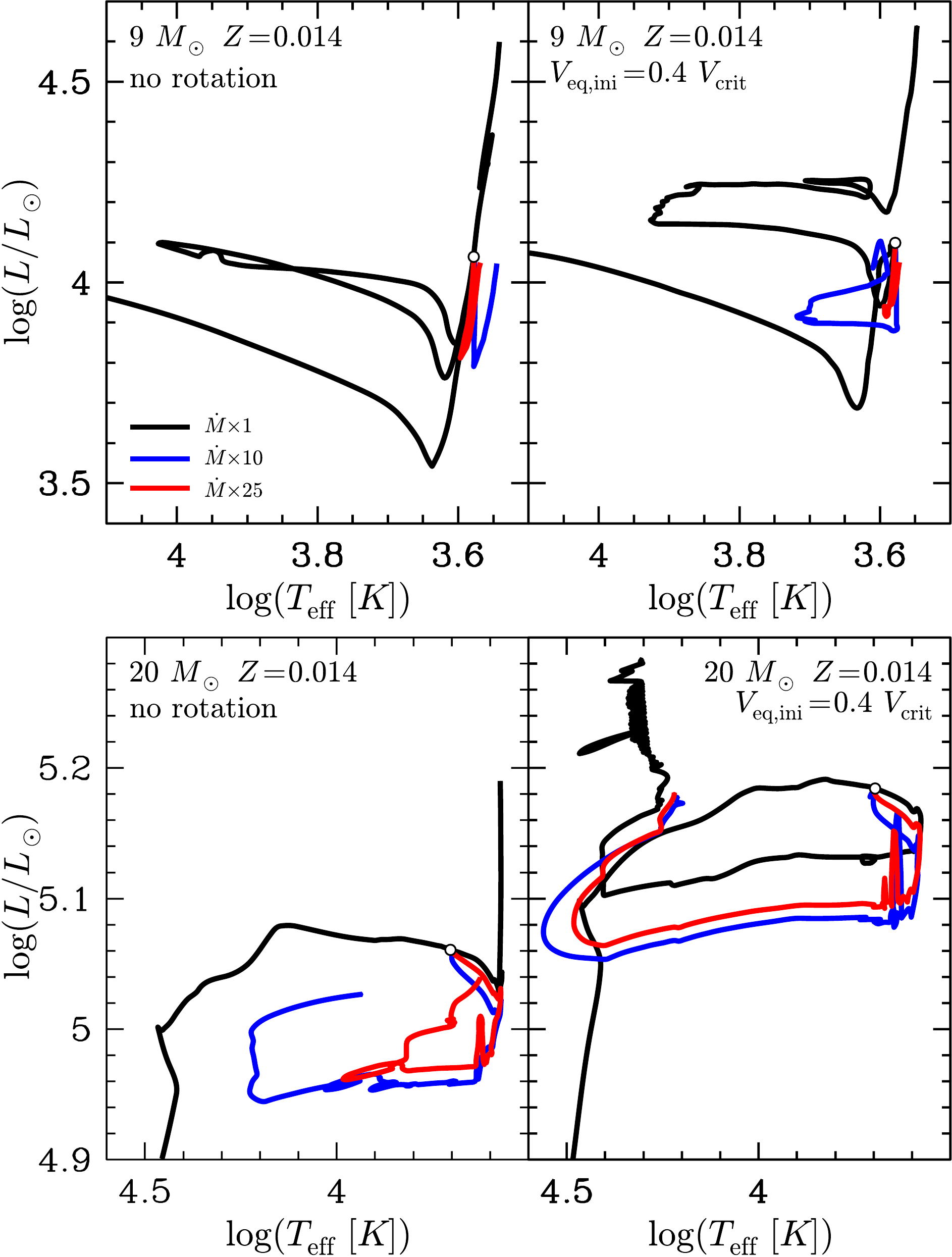}
   \caption{Evolutionary tracks for the 9 and 20 M$_\odot$ stellar models for various
   prescriptions of the RSG mass-loss rates. Only partial tracks are shown. In the upper panel, the lowest line shows the first crossing from blue to red
   of the HR gap. The beginning of the portion of the track computed with an enhanced RSG mass-loss rate is indicated by a small empty circle.}
      \label{Fig6HR}
\end{figure}
% macro RSG/fig6HR
% launching directory /Users/vega/Documents/latexAA/RSGXMdot/PAPIER

\begin{figure}[h]
\centering
\includegraphics[width=.39\textwidth]{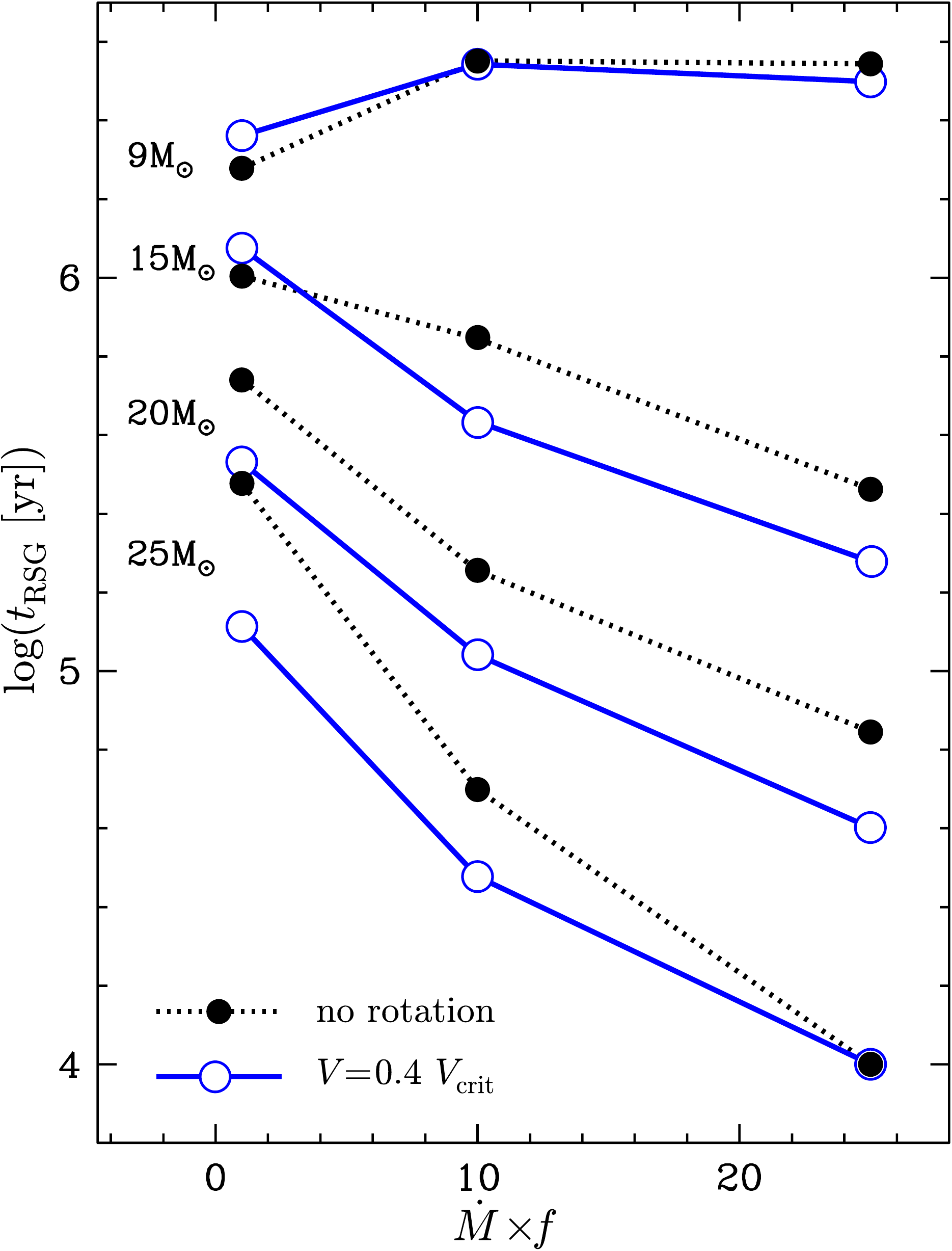}
\includegraphics[width=.39\textwidth]{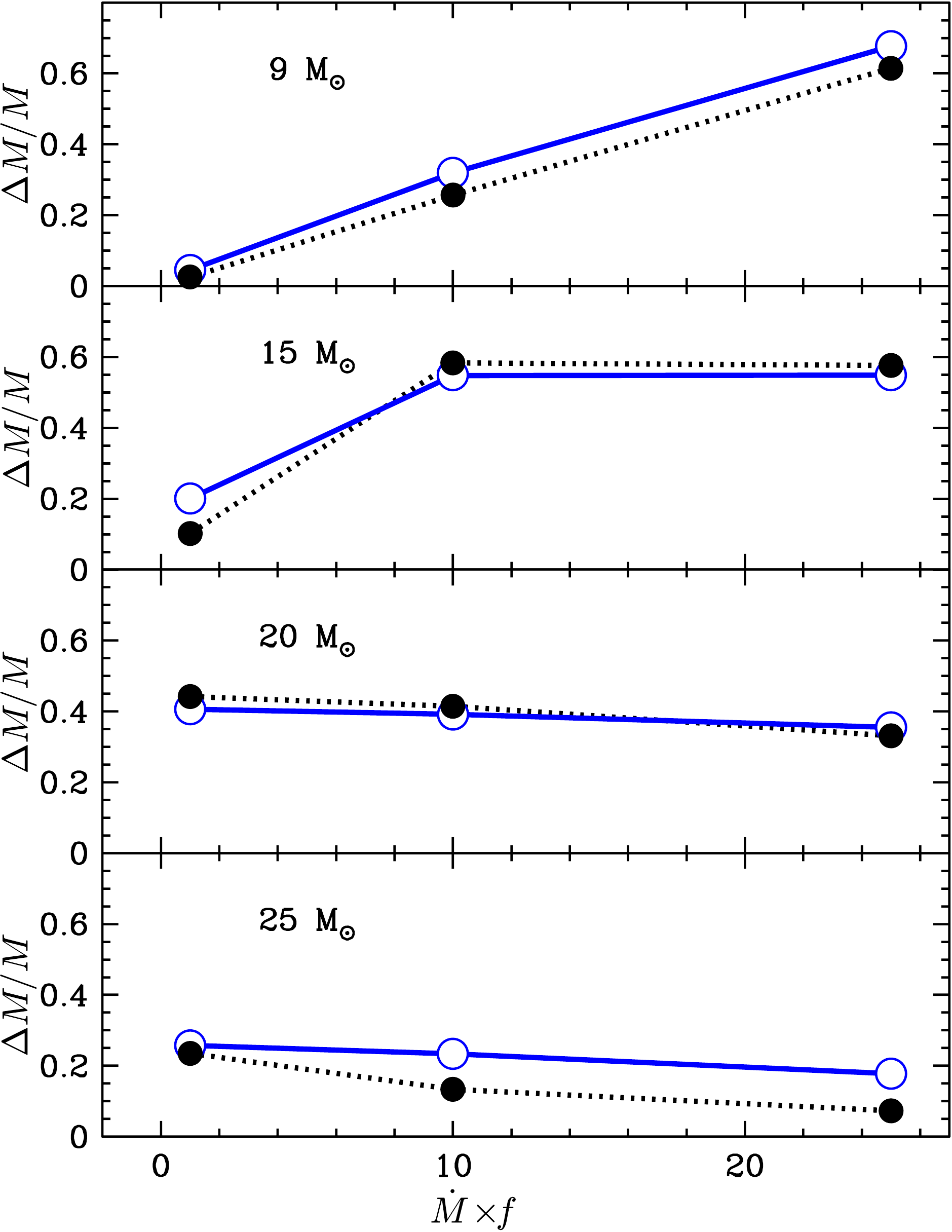}
   \caption{{\it Upper panel:} Logarithm of the red supergiant lifetimes as a function of the enhancement factor for the mass-loss rate during the red supergiant phase.  
      {\it Lower panel:} Fraction of the initial mass lost during the RSG phase, $\Delta M/M$, as a function of the enhancement factor  of the RSG mass-loss rate (see text).}
      \label{Fig2TI}
\end{figure}

We can wonder whether these enhancement factors for the mass-loss rates  are compatible with spectroscopically determined mass-loss rates. 
In Fig.~\ref{fig1VL}, mass-loss rate determinations for red supergiant stars are shown as a function of their luminosity.
The low (blue) shaded area covers the region where
 the sample of stars examined by \citet[][see their Tables 1 and 2 ]{Mauron2011} are located, while the upper (red)  shaded area shows the region covered by the dust-enshrouded red supergiants
 studied by \citet{vanLoon2005}. 
 We can see that at a given luminosity, the scatter of the mass-loss rates is very high as already noted by \citet{Jura90} and \citet{Josselin00}. 
 At a given luminosity, the mass-loss rates can show values which can differ by more than two orders of magnitude! 
 
  Is this very large scatter real or due to
 uncertainties in the techniques used to infer the mass-loss rates? In this work we assume that a significant part of that scatter is real.
 % This question is difficult to answer, but one would expect that the presence of dust would favor higher mass-losses, thus
 %the fact that dust enshrouded RSG have stronger mass-losses is at least not surprising.
%These dust-enshrouded red supergiants may represent a particular phase among red supergiant stars.
%Do all RSG go through such a phase?
%When does this phase occur?  For which type of  RSG stars?  And for how long?
%These questions have at the moment no answer. This is why 
In Fig.~\ref{fig1VL}, we compare the 
observationally deduced mass-loss rates with
the time averaged mass-loss rates obtained in the present work during the red supergiant phase (see the heavy lines). The averaged mass-loss rates 
have been obtained by simply extracting from the models the total mass lost and the duration of the RSG phase
(see the numbers in Table 1). We estimated also for each initial mass model the time averaged luminosity during the RSG phase. 
We see that, 1) the RSG mass-loss rate for the reference models \citep[those of][]{Ekstrom2012}, more or less go through the middle of the distribution of the points, indicating that
this choice of the mass-loss rate may be a good one for representing the averaged evolution; 2)
the models with mass-loss rates enhanced by a factor 10 and even 25 remain in a domain of mass-loss rates compatible with the mass-loss rates deduced from
spectroscopy. 
%{Since it is unclear the relative contribution between quiescent and eruptive RSG mass-losses, our choice of enhancement factors for the RSG mass-loss rates do not appear unrealistic.}
%The mass-loss rates for these models may still be much lower than the maximum values inferred from spectroscopy. Thus, these enhanced mass-loss
%rate models may be representative of either on these two cases: 1) stars which would lose mass at a regular rate equal to the one considered during their whole RSG phase or 2) undergoing strong short outbursts
%where the instantaneous mass-loss rate would still be much larger than the one considered here. 

%Likely this second situation is probably the one realized in nature. Indeed, some red supergiants show evidences of outburst events which may
%remove significant amount of mass in a short timescale. 

%%%%%%%%%%%%%%%////////////////////////////
 % macro RSG/fig1VL
% lauching directory /Users/vega/Documents/latexAA/RSGXMdot/PAPIER/PAPER

% macro RSG/fig2TI
%% lauching directory /Users/vega/Documents/latexAA/RSGXMdot/PAPIER/PAPER

\section{Evolutionary tracks and RSG lifetimes \label{evolution}}

The impact of the changes of the mass-loss rates during the red supergiant phase on 
the evolutionary tracks of the 9 and 20 M$_\odot$ models are shown in Fig~\ref{Fig6HR}.
One observes that in the case of the 9 M$_\odot$ model, the enhancement of the red supergiant mass-loss rate suppresses the blue loop, 
hence  increasing the time spent in the red supergiant phase, while for the 20 M$_\odot$, the mass-loss rate enhancement has the opposite effect. 
Rotation does not change qualitatively these trends \footnote{Actually, rotation produces marginal changes on the red supergiant lifetimes for the 9 M$_\odot$ stellar model (see Fig.~\ref{Fig2TI}), while it reduces the RSG lifetime by a little more than 0.2 dex for the 20 M$_\odot$ model. This is mainly due to the fact that the rotating models have higher luminosities, therefore suffer stronger mass-losses which reduce the RSG lifetimes.}.

Why does an increase in the mass-loss rate of the 9 M$_\odot$ reduce or even suppress the blue loop, while for the 20 M$_\odot$, on the contrary, it
favors a blue ward evolution? 
The physics of the blue ward evolution is not the same in the 9 and the 20 M$_\odot$ model. In the case of the 20 M$_\odot$, what makes the star evolve to the blue when the mass-loss increases
is the fact that the star becomes more and more homogenous, the helium core representing a still larger part of the total mass of the star \citep{Gia1967}. 
Typically, when the mass fraction of the He-core becomes greater than about 60-70\% of the total actual mass (this limiting fraction depends on the initial mass of the model considered), the star evolves back to the blue part of the HR diagram.

For the 9 M$_\odot$, we have a different situation. What makes the star to evolve back to the blue is the fact that, at some stage during the core helium burning phase, the core expands, implying by mirror effect,
a contraction of the envelope. An expansion of the core occurs more easily in a model which has a not too much massive core. 
%For the 15 M$_\odot$ models with standard mass-loss rates and above, the cores are too massive
%and core expansion cannot occur. This is why these models remains in the red part of the HRD and have no blue loop. For the 9 M$_\odot$, in contrast, this expansion can occur.
\citet{Lauterborn71} have shown that for such stars, the red-blue motion in the HR diagram
mainly depends on the gravitational potential of the He-core, $\Phi_{\rm core}$ and how it compares with some critical potential $\Phi_{\rm crit}(M)$ which grows with the stellar mass. One has that
when the gravitational potential of the He-core is greater than this critical potential, then the star remains in the RSG stage, while, when it is smaller, the core can expand and the envelope contracts, and the star
reaches a blue location in the HR diagram. So we can write
$$\Phi_{\rm core} > \Phi_{\rm crit}(M)\ \ \     {\rm Hayashi\  line},$$
$$\Phi_{\rm core} < \Phi_{\rm crit}(M)\ \ \       {\rm blue\  location}.$$
For masses equal or above 15 M$_\odot$, we are in the first situation,  for the 9 M$_\odot$, in the second one.

%This explains why when moderate mass loss rates are used as the ones used in  \citet{Ekstrom2012}, the 15 and 20 M$_\odot$ remain in the red, while the 9 M$_\odot$ shows a blue loop.
When the mass loss rates increase, for the 9 M$_\odot$ model, this decreases $\Phi_{\rm crit}(M)$. This favors the case where 
$\Phi_{\rm core} >  \Phi_{\rm crit}(M)$, and an evolution keeping the star in the red part of the HR diagram. For the more massive stars, an increase of the mass loss rates also disfavors the evolution back to the blue
due to the mirror effect. But as explained above, these stars may nevertheless evolve back to the blue because due to their massive convective cores, when mass loss is strong, the structure of the star becomes
more and more homogeneous and thus these stars
evolve in the direction of the helium-rich homogeneous sequence in the HRD, {\it i.e} in the blue part of the HR diagram.

\begin{table*}
\caption{Core He-burning lifetimes, red supergiant lifetimes, time-averaged mass-loss rates during the red supergiant phase as well as different properties of the last computed models. The abundances are in mass fractions. The metallicity is solar.}
\label{Tab1}
\centering
\begin{tabular}{r r r r r r r r r r r r}
\hline\hline
    $M_{\rm ini}$ & $\dot{M}$ & $t_{\rm He}$ & $t_{\rm RSG}$ & $<\dot{M}>_{\rm RSG}$ & $M_{\rm fin}$ & Log $L/L_\odot$  & Log $(T_{\rm eff}/K)$  & H$_{\rm surf}$ & He$_{\rm surf}$ & N/C$_{\rm surf}$ &  N/O$_{\rm surf}$ \\
    $M_\odot$ &  & Myr & Myr & $M_\odot$  $yr^{-1}$ & $M_\odot$ & & & & & & \\
 \hline
 & & & & & & & & & & & \\ 
 \multicolumn{12}{c}{$\upsilon_{\rm ini}$=0.} \\
 & & & & & & & & & & & \\      
    9   & 1$\times \dot{M}_{\rm stan.}$   &  3.6693 & 1.90 & 8.66e-08 & 8.7651   & 4.5955 & 3.5416 & 0.6064 & 0.3788 & 1.76      & 0.67  \\
    15 & 1$\times \dot{M}_{\rm stan.}$   &  1.3344 & 1.01 & 1.50e-06 & 13.1739 & 4.7964 & 3.5592 & 0.6763 & 0.3096 & 2.25      & 0.56  \\
    20 & 1$\times \dot{M}_{\rm stan.}$   &  0.8921 & 0.55 & 1.62e-05 & 8.6347   & 5.1823 & 3.5730 & 0.4755 & 0.5106 & 86.25    & 4.31  \\
    25 & 1$\times \dot{M}_{\rm stan.}$   &  0.7035 & 0.30 & 1.34e-05 & 8.2893   & 5.3790 & 4.4332 & 0.1558 & 0.8305 & 117.14  & 74.55  \\
 & & & & & & & & & & & \\    
    9   & 10$\times \dot{M}_{\rm stan.}$   &  3.5306 & 3.56 & 6.49e-07 & 6.6876   & 4.0496 & 3.5685 & 0.7033 & 0.2826 & 1.57      & 0.42  \\
    15 & 10$\times \dot{M}_{\rm stan.}$   &  1.4212 & 0.70 & 1.32e-05 & 4.8563   & 4.7789 & 3.6116 & 0.4409 & 0.5453 & 118.33   & 5.07  \\
    20 & 10$\times \dot{M}_{\rm stan.}$   &  0.8872 & 0.18 & 5.00e-05 & 6.6413   & 5.0392 & 3.6257 & 0.4653 & 0.5209 & 236.67   & 5.07  \\
    25 & 10$\times \dot{M}_{\rm stan.}$   &  0.6791 & 0.05 & 1.30e-04 & 8.2648   & 5.2209 & 4.2322 & 0.4469 & 0.5393 & 185.00  &  7.40  \\
 & & & & & & & & & & & \\ 
     9   & 25$\times \dot{M}_{\rm stan.}$  &  3.4566 & 3.50 & 1.58e-06 & 3.4715   & 4.0474 & 3.5448 & 0.7033 & 0.2826 & 1.57      & 0.42  \\
    15 & 25$\times \dot{M}_{\rm stan.}$   &  1.4056 & 0.29 & 3.04e-05 & 4.6011   & 4.7867 & 3.7388 & 0.4409 & 0.5453 & 118.33   & 5.07  \\
    20 & 25$\times \dot{M}_{\rm stan.}$   &  0.8798 & 0.07 & 1.27e-04 & 6.5014   & 5.0268 & 3.9331 & 0.4652 & 0.5210 & 236.67   & 5.07  \\
    25 & 25$\times \dot{M}_{\rm stan.}$   &  0.6867 & 0.01 & 2.56e-04 & 8.2843   & 5.2606 & 4.2330 & 0.4180 & 0.5683 & 130.00  &  13.45  \\
  & & & & & & & & & & & \\ 
 \multicolumn{12}{c}{$\upsilon_{\rm ini}$=0.4$\upsilon_{\rm crit}$} \\
 & & & & & & & & & & & \\      
    9   & 1$\times \dot{M}_{\rm stan.}$   &  3.8474 & 2.30 & 1.78e-07 & 8.5175   & 4.6309 & 3.5476 & 0.6487 & 0.3372 & 6.33     & 0.86  \\
    15 & 1$\times \dot{M}_{\rm stan.}$   &  1.3314 & 1.19 & 2.56e-06 & 11.5156 & 4.9588 & 3.5598 & 0.5951 & 0.3908 & 7.00      & 1.11  \\
    20 & 1$\times \dot{M}_{\rm stan.}$   &  0.8616 & 0.34 & 2.54e-05 & 7.1785   & 5.2811 & 4.3087 & 0.2381 & 0.7482 & 130.00    & 15.92  \\
    25 & 1$\times \dot{M}_{\rm stan.}$   &  0.6276 & 0.13 & 7.17e-05 & 9.6896   & 5.5032 & 4.3914 & 0.5120 & 0.9259 & 101.87   & 25.08 \\
 & & & & & & & & & & & \\    
    9   & 10$\times \dot{M}_{\rm stan.}$   &  3.4742 & 3.49 & 8.25e-07 & 6.1108   & 4.0500 & 3.5726 & 0.6539 & 0.3320 & 5.64     & 0.83  \\
    15 & 10$\times \dot{M}_{\rm stan.}$   &  1.4133 & 0.43 & 1.96e-05 & 5.3064   & 4.9444 & 3.6754 & 0.3916 & 0.5945 & 81.25   & 3.10  \\
    20 & 10$\times \dot{M}_{\rm stan.}$   &  0.8539 & 0.11 & 8.70e-05 & 7.2971   & 5.1810 & 4.2195 & 0.3780 & 0.6081 & 170.00   & 4.00  \\
    25 & 10$\times \dot{M}_{\rm stan.}$   &  0.6150 & 0.03 & 1.95e-04 & 9.7153   & 5.3717 & 4.5014 & 0.3186 & 0.6676 & 182.50  &  6.64  \\
 & & & & & & & & & & & \\ 
     9   & 25$\times \dot{M}_{\rm stan.}$  &  3.3787 & 3.15 & 1.93e-06 & 2.5951   & 4.0575 & 3.6062 & 0.6524 & 0.3325 & 5.70    & 0.83 \\
    15 & 25$\times \dot{M}_{\rm stan.}$   &  1.2443 & 0.19 & 4.57e-05 & 5.3091   & 4.9288 & 3.6213 & 0.3891 & 0.5970 & 64.00   & 3.05 \\
    20 & 25$\times \dot{M}_{\rm stan.}$   &  0.8542 & 0.04 & 2.37e-04 & 7.1887   & 5.1786 & 4.2109 & 0.3755 & 0.6106 & 170.00   & 4.00  \\
    25 & 25$\times \dot{M}_{\rm stan.}$   &  0.6207 & 0.01 & 4.45e-04 & 9.6199   & 5.3559 & 4.5549 & 0.3298 & 0.6564 & 179.49  &  5.00  \\
& & & & & & & & & & & \\      
 \hline
\end{tabular}
\end{table*}

Various properties of the present stellar models are indicated in Table 1. 
Comparing the outputs of the stellar models obtained with different RSG mass-loss rates, one deduces the following points:
\begin{itemize}
\item The core He-burning lifetimes are little affected by a change of the RSG mass-loss rates. This comes from the fact that
the stellar winds are never strong enough to modify significantly the He-core masses.
\item The RSG lifetimes are strongly affected by a change of the RSG mass-loss rates. This can be seen also in the upper panel of Fig.~\ref{Fig2TI}.
We see that
when the red supergiant mass-loss rate increases, the duration of the red supergiant lifetime increases for the 9 M$_\odot$ (by a factor 2 
when the mass-loss rate is increased by a factor 10 with respect to the standard value), as the result of the blue loop disappearance.
In contrast, the RSG lifetimes
decrease for the 20 M$_\odot$ model computed with enhanced RSG mass-loss rates: 
by a factor 4 for the same change of the mass-loss rates as for the 9 M$_\odot$ model. 
This comes from the fact that the 20 M$_\odot$ models evolve back to the blue part of the HR diagram when they undergo strong mass-losses
\citep[see also][]{Salasnich1999, Vanbeveren2007, Yoon2010, Georgy2012,gme13,Groh2013}.
Still increasing the mass-loss rates, the duration of the red supergiant phase for the 9 M$_\odot$ does no longer much change, while for the
20 M$_\odot$, it continues to decrease, by about a factor 2.5 passing from 10 $\times$ to 25 $\times$
the standard mass-loss rate.
\item The lower panel of Fig.~\ref{Fig2TI} shows the fraction of the initial mass which is lost by stellar winds during the RSG phase when various mass-loss prescriptions are used. 
Let us first comment the 15 M$_\odot$ models. We see that with the standard mass-loss rate, 10-20\% (depending on rotation) is lost during the RSG phase. When the mass-loss rates are increased by an order of magnitude, the fraction increases up to 55-60\%. Increasing the mass-loss rates further (from 10x to 25x) does not change this fraction further. This saturation of the mass lost comes from the fact that 
a star leaves the RSG region when the envelope mass is reduced to a certain value for a given core mass.
As a consequence,  an increase of the mass-loss rate reduces the time spent in the RSG phase but not the total mass lost.
The total mass lost during the RSG phase remains around 40\% in the case of the 20 M$_\odot$ stellar model and below 30\% for the 25 M$_\odot$ stellar model.
{\it So for stars with initial masses between 15 and 25 M$_\odot$, a blue ward evolution implies the loss of a relatively well fixed amount of mass}.
The mass lost  is not sufficient for producing a WC star or said in other words a naked CO core (WC are Wolf-Rayet stars characterized by strong carbon and oxygen emission lines). 
%Indeed the CO core masses (see Table 1) are not uncovered by the removal of part of the envelope during the RSG phase.
%For the 15, 20 and 25 M$_\odot$ models, this amounts to 60\% (mass lost equals to 9 M$_\odot$), 40\% (8 M$_\odot$) and around 15\% (3.75 M$_\odot$), respectively, while their CO-core masses is at maximum of  2.78, 4.73 and 7.09 M$_\odot$, respectively (see Table 1).  The CO-core masses are significantly smaller than respectively 15-9=6 M$_\odot$, 20-8=12 M$_\odot$ and 25-3.75=21.25 M$_\odot$, which is the amount of mass that has to be lost to produce a WC star. 
Some mass can be lost after the RSG stage, but this amount is quite modest assuming only quiescent winds. Since the stars become Luminous Blue Variabes (LBVs) when they come back to the blue after losing significant amounts of mass as a RSG \citep{gme13}, additional mass could be lost in LBV eruptions, but the total mass lost is unclear.  So already at that stage, one can give an answer to one of the question raised in the introduction: Can such a bluewards evolution (induced by an enhanced RSG mass-loss rate) explain the low luminosity WC stars \citep{Georgy2012WR}? The answer seems to be no, confirming the results from  \citet{Georgy2012}.
This conclusion remains valid in case the mass-loss would be due to a mass transfer episode to a binary partner  in a close multiple system
during the red supergiant phase rather than to an increase of the mass-loss rate as computed here. 
It is interesting here to mention also the results by \citet{Chieffi2013} who obtained for the 20 M$_\odot$ an evolution back to the blue after a RSG phase. The mass loss rates used by these authors somewhat differ from the standard one used in the present model. Their 20 M$_\odot$ enters the WR phase but never becomes a WC star somewhat supporting the present conclusion.
\item In general, including the effects of rotation does not change much the results.
For a given initial mass, it lowers the RSG lifetimes, while it does not much affect the total mass lost during the RSG phase.
This is due to the fact that,  when the effects of rotation are included, for a given initial mass, the RSG is more luminous, thus the mass-losses are stronger.
As explained above, this shortens the RSG lifetimes.
\end{itemize}

\section{Properties of red supergiants}

Figure~\ref{hrrsgwr} shows the location of the Galactic RSGs observed by \citet{Levesque2005} together with the evolutionary tracks computed with and without rotation and for different RSG mass-loss rates.
We see that changing the RSG mass-loss rate has no strong impact on the effective temperatures of RSGs, but it modifies the range of luminosities covered by a given initial mass when it is a RSG.
The models with an enhanced RSG mass-loss rate
encompass a smaller range of luminosities during the RSG phase than those with standard mass-loss rate (compare for instance the RSG part of the tracks for the 15 M$_\odot$ models). 
This is caused by the much shorter lifetimes spent in the RSG phase for the enhanced mass-loss rate models. 
The use of tracks computed with different RSG mass loss rate to determine the range of initial masses of an observed RSG, provide somewhat different values.
To give a numerical example, if we observe a red supergiant with a luminosity equal to Log $L/L_\odot=5.0$, the standard mass loss rate
can produce such a red supergiant starting for instance from a 15 $M_\odot$ model, while the 10 $\times$ and 25 $\times$ models require a higher initial mass, around 18-19 $M_\odot$.
%However, this is difficult to be constrained by observations since the initial mass of stars is a quantity which is not known precisely.

The changes of RSG lifetimes shown in Fig.~\ref{Fig2TI} may have an observable impact on the luminosity function of RSGs in regions of constant star-formation rate\footnote{In coeval stellar populations, the range of masses of stars, which at a given age, are red supergiants is likely too small, to allow such an effect to be visible.}. 
To assess in a quantitative way, a population synthesis model is needed. This will be done in a future work.
At the moment,  we may already have an idea of the importance of the effect using a more simple approach.
In the left panel of Fig.~\ref{lumlf}, we have plotted the  times spent in the RSG stage for our different initial mass models weighted by the Salpeter's initial mass function. We use the quantities for the 9 M$_\odot$ as normalization. 
More precisely, the vertical axis of the left panel of Fig.~\ref{lumlf} indicates $\lg (t_{\rm RSG}(M_{\rm ini})/t_{\rm RSG}(9M_\odot) \times M_{\rm ini}^{-2.35}/9^{-2.35})$, where $t_{\rm RSG}(M_{\rm ini})$ is the RSG lifetime of a stellar model with an initial mass  $M_{\rm ini}$.
To each initial mass we have attributed a time-averaged luminosity during the RSG phase as it can be deduced from the evolutionary tracks (these average luminosities are log $L$/L$_\odot$ $\approx$ 3.9, 4.7, 5.0 and 5.3 for respectively the 9, 15, 20 and 25 M$_\odot$).  We see that when the RSG mass-loss rate increases, the slope of this "luminosity function"  becomes steeper. When the luminosity
increases from Log $L/{\rm L}_\odot$ equal 4 to 5, the number of star decreases by a factor 30 when standard mass-loss rates are used, while it decreases by a factor 500 when models with 25 $\times$ the standard mass-loss rate are used. From the left panel of Fig.~\ref{lumlf}, one deduces that the impact of a change of mass-loss is much stronger than the impact due to rotation (at least for the ranges of values explored here). 

To use such a feature to constrain the stellar models, a few conditions must be fulfilled: 1) it requires some completeness of the sample along the whole RSG sequence; 
2) the red supergiants, originating from masses equal or larger than 9 M$_\odot$ should be distinguished from luminous AGB stars; 
%3)  the bulk of the RSG stars must undergo some well defined RSG mass-loss rates
3) the star formation rate should have been constant in the last 30 million years or so. 
In case these conditions are fulfilled then, the use of the RSG luminosity function to constrain the models  has two interesting advantages : 1) it is based on a very simple observations (number ratios);  2) this quantity will provide some constraint on the averaged mass-loss during the RSG phase which is a more useful quantity  than the instantaneous mass-loss rate which actually might not be the one responsible for the loss of most of the material during that phase. 

\subsection{Effective temperatures and radii of RSG}

\begin{figure*}
%: fig TcrcZ014_Mcc_rot.eps
\centering
\includegraphics[width=.94\textwidth, angle =0]{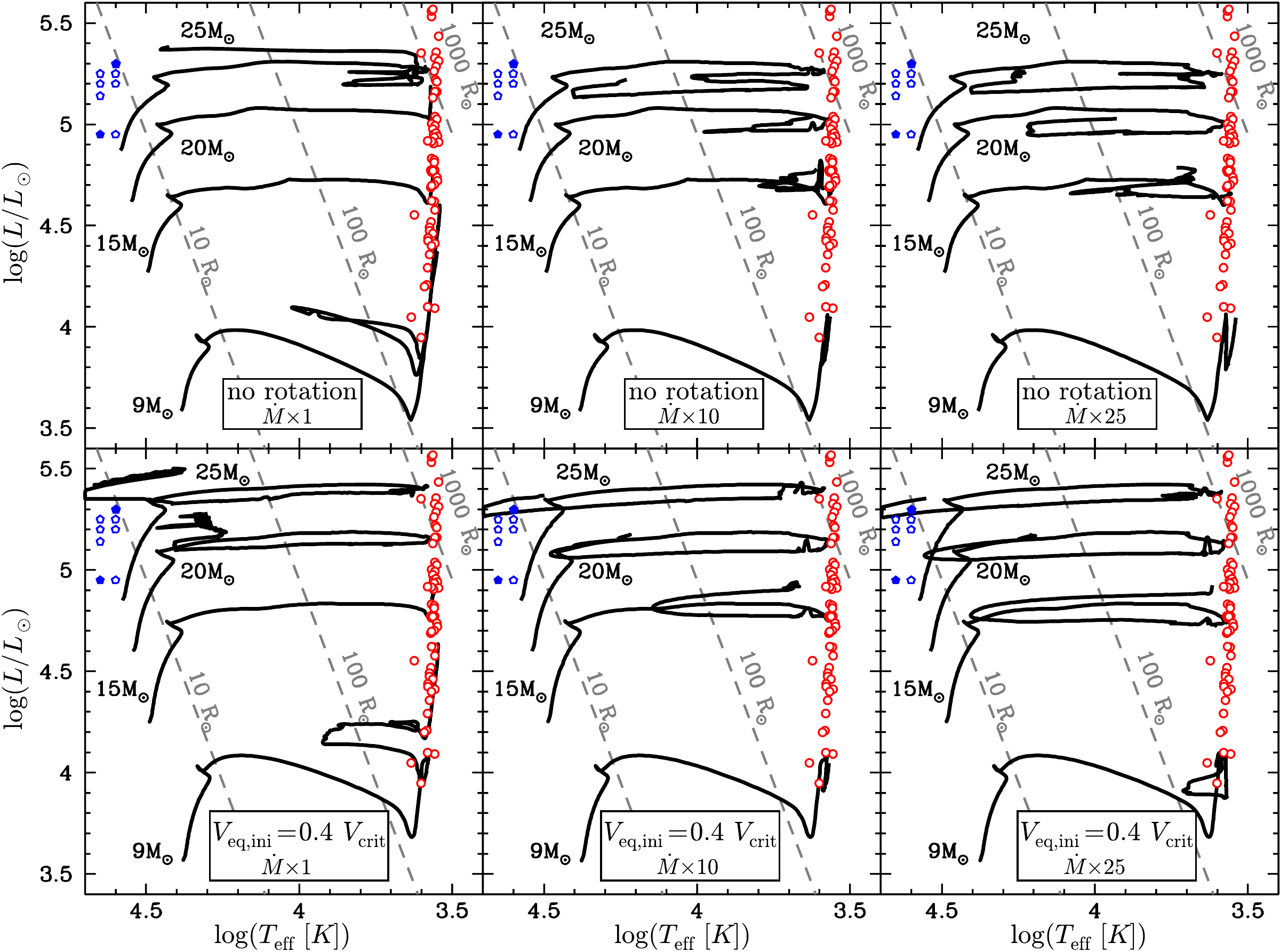}

   \caption{Evolutionary tracks of models with different initial velocities (non rotating on the top, rotating on the bottom) with different RSG mass-loss rates (increasing from left to right).
   The (red) empty circles are the positions of the galactic red supergiants observed by \citet{Levesque2005}. The blue pentagons
   are the positions of WC stars as observed by \citet{Sander12}.
}
\label{hrrsgwr}
\end{figure*}

%macro RSG, rayL, a activer dans latexAA/RSGXMdot/PAPIER, after mv dans PAPER/.

The effective temperature or  the radius of a RSG is not a very strong prediction of the stellar models since it depends on the choice of the mixing length used to compute the non-adiabatic convection in the envelope.
Also the modeling of the convection may be complicated by the fact that the velocity of the convective cells may approach or even supersedes the sound speed implying shocks \citep[see e.g.][]{BOOK2009}. 
All the present models assume a mixing length, 1.6 times the pressure scale height \citep[see Section 2, note that in these models we did not use a mixing length proportional to the density scale height as for instance was done in][]{MaeMey1987}. The value 1.6 is calibrated on the Sun. It happens that this mixing length value provides also a good match of the observed positions of the red giant  and supergiant branch 
\citep[see Fig.~2 in][and Fig.~\ref{hrrsgwr}]{Ekstrom2012}.

\begin{figure*}
%: fig TcrcZ014_Mcc_rot.eps
\centering
\includegraphics[width=.405\textwidth]{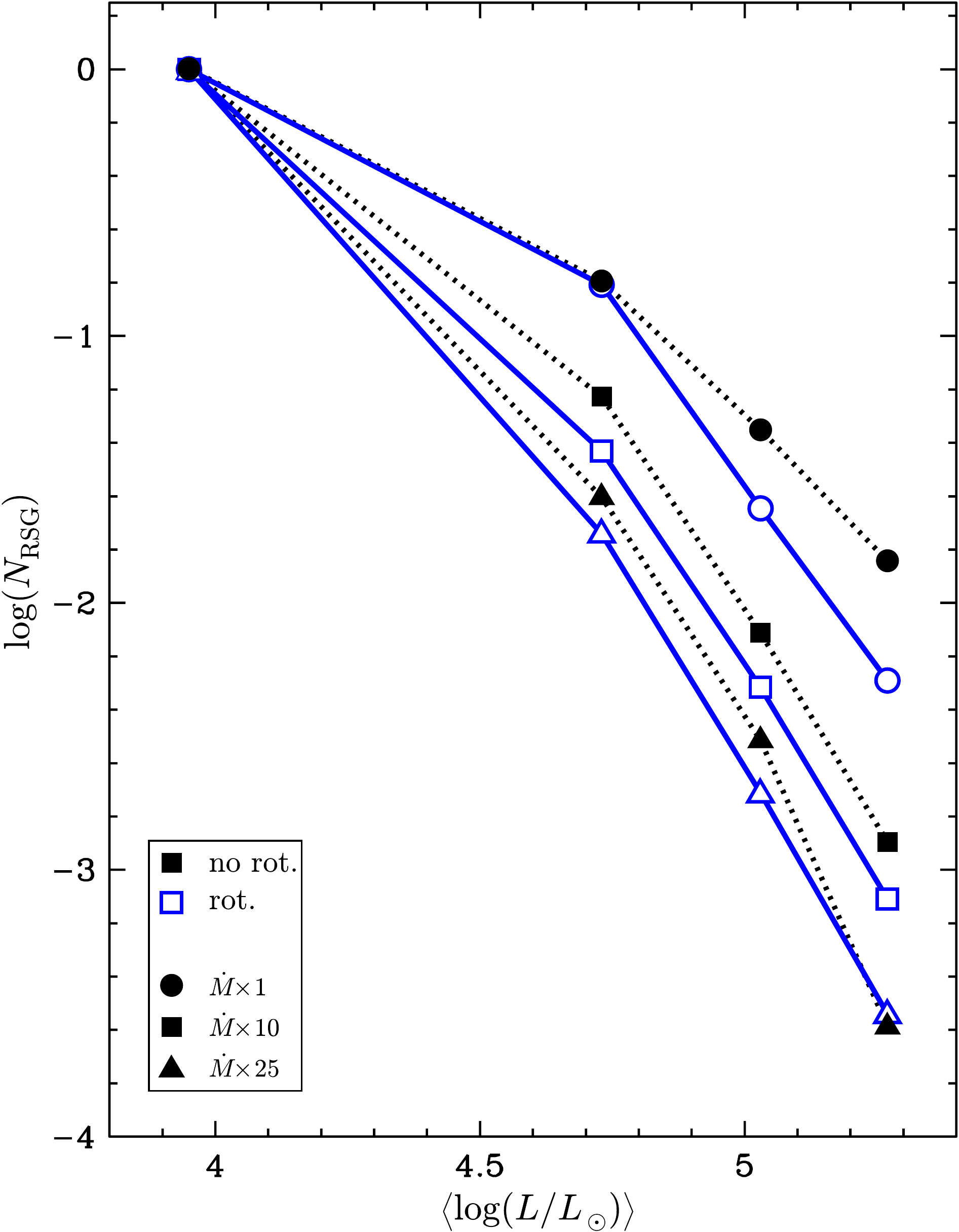}\hskip 2cm\includegraphics[width=.40\textwidth]{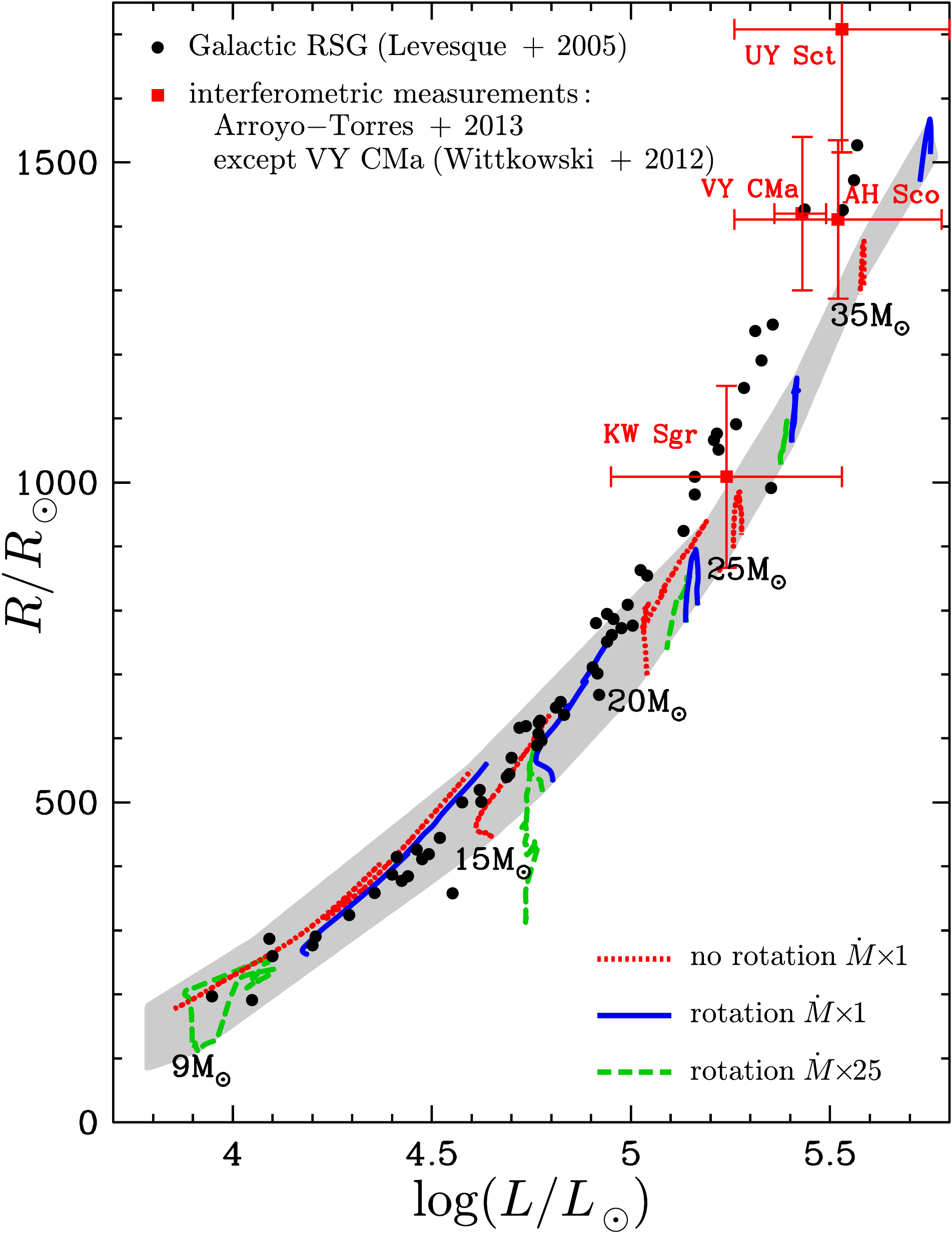}
   \caption{{\it Left panel:} Number of red supergiants as a function of luminosity in a constant star formation rate region (see text).
%   Red supergiant lifetimes weighted by a Salpeter IMF as a function of the time-averaged luminosity during the RSG phase (see text).
Numbers are normalized to the 9 M$_\odot$. Dots indicate the values obtained for the 9, 15, 20 and 25 M$_\odot$ models from left to right.
   {\it Right panel:} Radii of various stellar models during the RSG phase (Log T$_{\rm eff} < 3.6$) as a function of the luminosity. 
   The shaded area covers the region covered by the models with different initial masses, rotations and mass-loss rates. The 25 times enhanced, rotating 15 M$_\odot$ is the only model
   which evolves out of this region. 
  The initial mass is indicated at the minimum luminosity predicted by the models for this mass.}
      \label{lumlf}
\end{figure*}

The radii of the present stellar models during the RSG phase are plotted as a function of the luminosity in the right panel Fig.~\ref{lumlf}. 
%We defined the RSG phase as the phase during which the effective temperature is strictly below 3.6.
We see that the radii of the models span a large range of values going from 100 up to 1570 R$_\odot$. 
During the RSG phase, the variation of the radii for a given initial mass can be quite large for the 9 and 15 M$_\odot$ (factor 6 for the 9 M$_\odot$
and factor 2 for the 15 M$_\odot$). The range of radii covered by the more massive models is much smaller (5-20\%). 
At a fixed luminosity, the range of possible radii for the RSG remains modest. Typically, the shaded area in the right panel of Fig.~\ref{lumlf} has a width of 50-100 R$_\odot$.

In the radius-luminosity plane, rotation pushes the tracks of a given mass at higher luminosities but along the same general sequence as the one described by the non-rotating models.
Enhancing the mass-loss rates during the RSG phase produces very small effects in general in this diagram. It shifts somewhat the tracks to slightly lower luminosities but again along
the same general trend defined by the standard models. There is one exception however in the case of the rotating 15 M$_\odot$ with 25 times the standard mass-loss rates. This model extends
much below the general trend. This comes from the fact that the luminosity during the RSG phase decreases a lot for this model as a result of the mass-losses. 

When comparisons are made with observed values (see the right panel of Fig.~\ref{lumlf}),
we note a good agreement between models and observations for luminosities between 3.9 up to 5.1-5.2.
For higher luminosities, the radii given by the models are too small by about 10-20\%  with 
respect to the observations. On this plot some interferometric determinations are also indicated
and are also larger than the predicted values. A possibility on the side of the theoretical models to improve the situation, namely to
produce larger radii at higher luminosities, would be to decrease the mixing length in the upper luminosity range.
%The question then would be what would be the physical reason explaining this change of mixing length at higher luminosity. 
It might be also that the values inferred from the observations are slightly too high. The radii of red supergiants is not an easy quantity to measure, since these stars have a very extended and tenuous atmosphere. One can thus easily understand that, depending on wavelength, the radius may differ from the way the radius is determined in our stellar models. Moreover in case the wind is strong enough, it can be optically thick. In that case the surface is no longer observable and a pseudo photosphere at larger radius appears! 

Some authors find that the observed radii of RSGs are smaller than those presented in the right panel of Fig.~\ref{lumlf}. The arguments come from two different approaches:
1) \citet{Davies2013} recently redetermined the effective temperatures of RSGs in the Magellanic Clouds finding warmer temperatures than found by \citet{Levesque2006} and thus favoring radii of red supergiants about
20-30\% more compact. This affects all radii  in the whole range of luminosities. The metallicity however
is different than solar, therefore we can wonder whether applying their techniques on solar metallicity RSGs, these authors
would obtain a similar systematic difference. We note that the \citet{Levesque2006}  effective
temperatures of Galactic RSGs agree very well with results obtained quite independently by interferometric techniques  \citep[see Table 2 in][]{vanBelle2009}.
2) Another indirect argument pointing towards smaller radii for red supergiants is the one by \citet{Dessart2013} who argued, based on properties of the type II-P supernova light 
curve that the progenitors should have a much more smaller radius than commonly assumed (note that this
constraint applies to RSG which are the end point of the evolution of the considered star and not to RSGs in general).  More precisely, \citet{Dessart2013} show that light curves arising from the explosion of more compact RSG have less blue and shorter plateaus in better agreement with the observation. Typically, these authors show that their 15 M$_\odot$ model with a radius of 500 R$_\odot$ much better reproduces the light curves observed in different filters for the type II-P SN 1999em. Looking at the right panel of Fig.~\ref{lumlf}, one sees that the 15M$_\odot$ model ends with radii between 300 and 800 R$_\odot$ depending on the RSG mass-loss rate used. In order to obtain, 
at the end of the evolution, a radius of 500 R$_\odot$, an enhancement factor of the mass-loss rate during the RSG phase between 10 and 25 should be used. Note however, that this is not the only way to obtain a more compact
radius. Another solution would be to change the way to compute the outer convective zone \citep{MaeMey1987, Dessart2013}.

The above discussion indicates that the radii (or effective temperatures) of RSGs are still in doubt.  
At the moment, we conclude that the mixing length considered in the present models do in general a good job. An enhanced
RSG mass-loss rate can explain some more compact radii for RSGs at the pre-supernova stage, in the mass range around 15 M$_\odot$. 

%So, on one side, the radii of RSG seem to be underestimated for luminosities higher than about 5.1-5.2 (see Fig.~\ref{hrrsgwr}), and, on the other side, to be overestimated with respect
%to some new determinations of the effective temperatures of RSG. Since, as explained above, the radii of the effective temperatures of RSG is quite sensitive to the choice of the mixing length parameter,
%the theory at the moment cannot decide which, among the two situations, is the one the nearest to reality. We can however at the moment retain the following general points: 
%1) this problem of the radii/effective temperatures of red supergiants do affect stellar evolution. Depending whether star reach a more or less red positions during that phase, the mass-loss rates will be different
%(less mass is lost likely when the star is more compact), and thus it is important to improve our knowledge\footnote{Note the in the present work, since the mass-loss rate
%is considered as a free parameters, the question of which effective temperature/radius is realized is not so much important.}; 2) The constraints on the radii of red supergiants cannot be used to discard any of the models presented %here, even those computed with the most extreme mass-loss rates during the RSG phase. 

%void table 1 du papier de UTROBIN (placer les 7 IIP progenitors in HR diagramme and see whether it is fitted by models of the correct mass).
%Antares (Ohnaka et al. 2013) OK with tracks. 

%radii

% \begin{figure}
%\centering
%\includegraphics[width=.40\textwidth]{rayL.pdf}
%   \caption{Radii of various stellar models during the RSG phase (Log T$_{\rm eff} < 3.6$) as a function of the luminosity. Short-dashed (red) lines correspond to non-rotating,
%   standard mass-loss rate models, the continuous (blue) and long-dashed lines (magenta) are for rotating models computed with standard and 25 times mass-loss rates during the RSG phase.
%   The shaded area covers the region covered by the models with different initial masses, rotations and mass-loss rates. The 25 times enhanced, rotating 15 M$_\odot$ is the only model
%   which evolves out of this region. The black points corresponds to the observed galactic sample of RSG by \citet{Levesque2005}. The points with error bars correspond to interferometric measures by
%  \citet{Wittkowski12} for VY CM, and \citet{Arroyo13} for AH Sco, UY Sct and KW Sgr.
%  The initial mass is indicated at the minimum luminosity predicted by the models for this mass.}
%      \label{rl}
%\end{figure}

\subsection{The surface composition}

\begin{figure*}
%: fig TcrcZ014_Mcc_rot.eps
\centering
\includegraphics[width=.47\textwidth, angle =0]{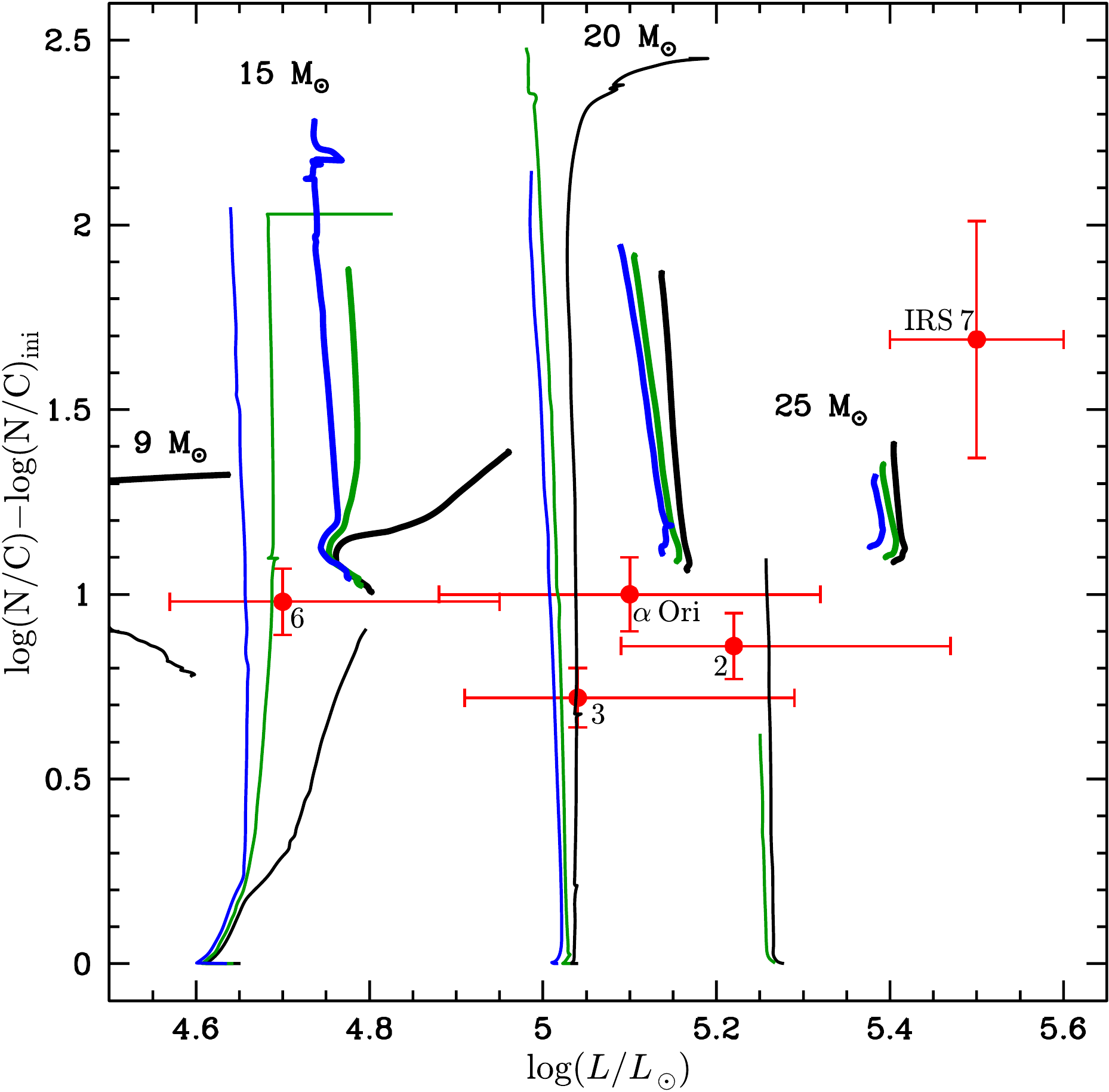}\includegraphics[width=.47\textwidth, angle =0]{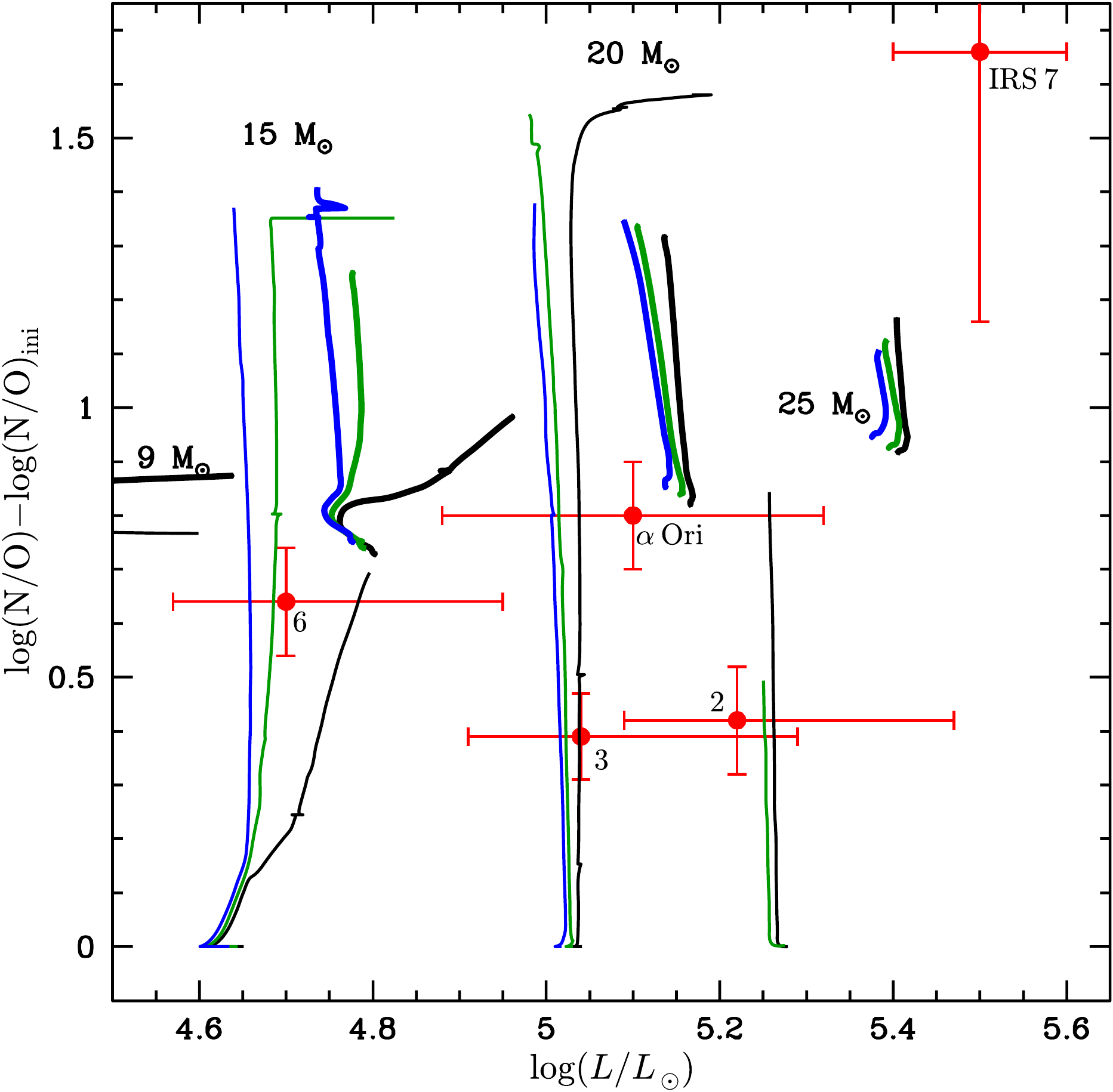}
\includegraphics[width=.47\textwidth, angle =0]{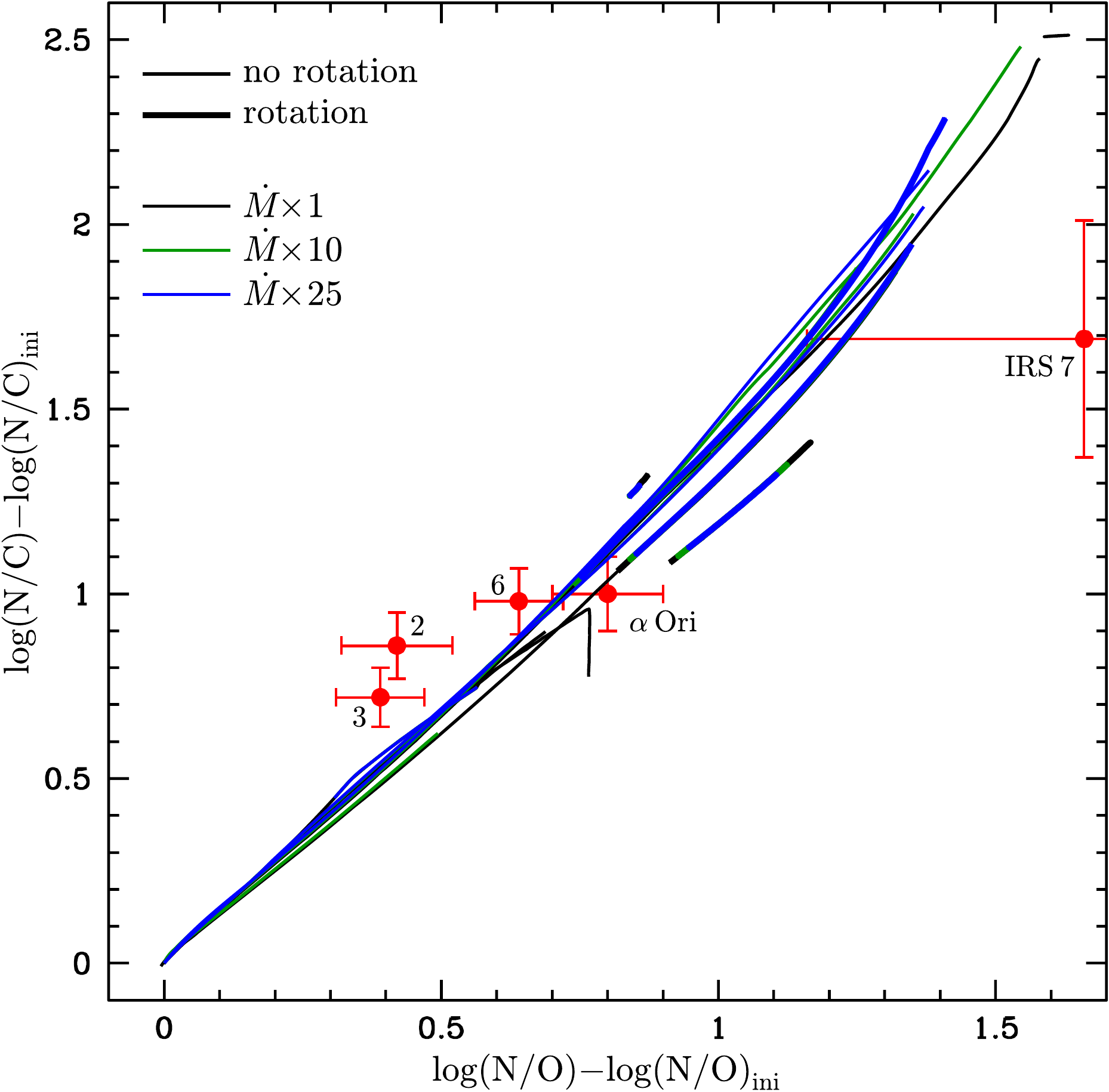}\includegraphics[width=.47\textwidth, angle =0]{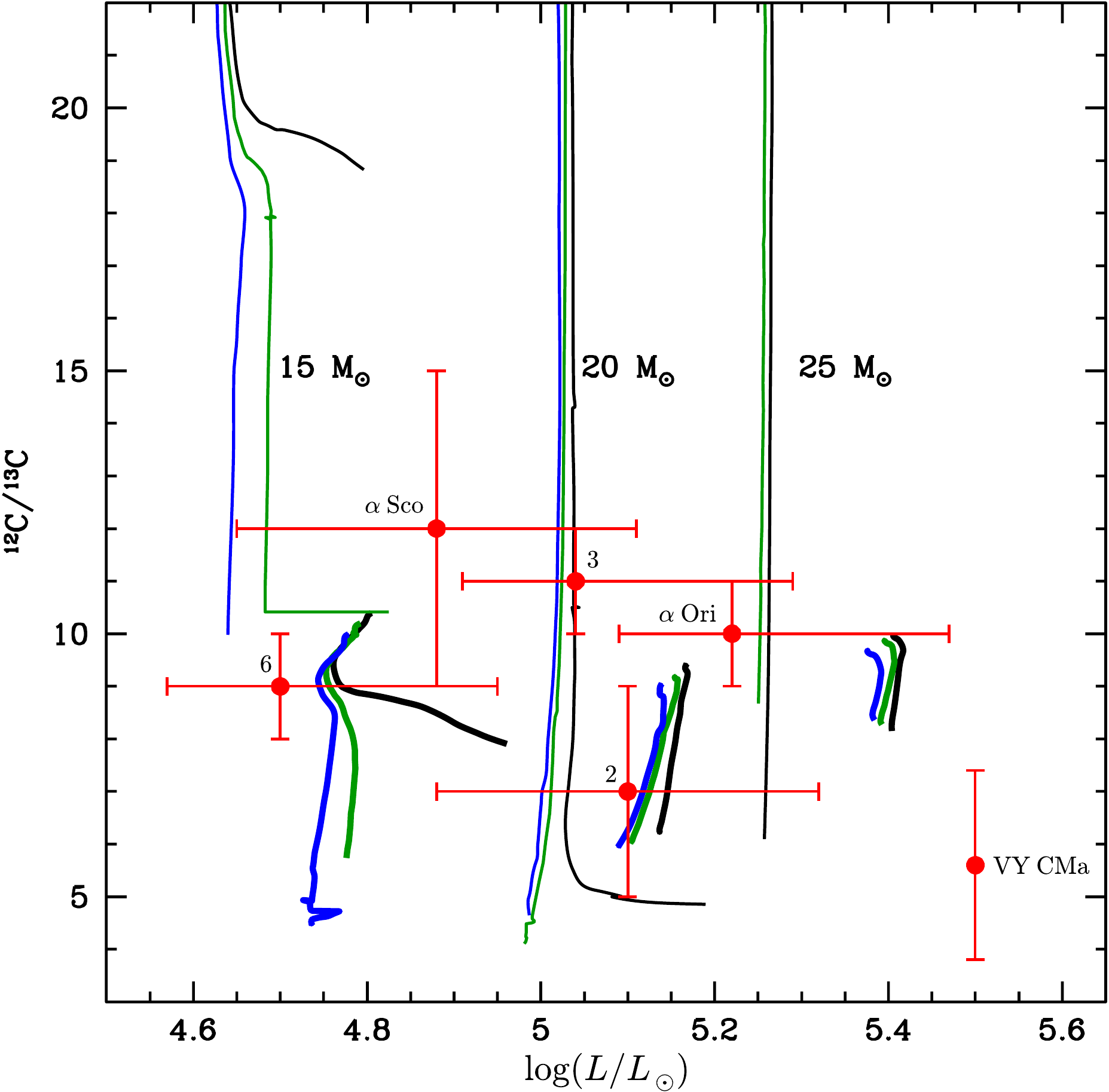}

   \caption{
   {\it Upper left panel:} Surface N/C ratios during the red supergiant phase normalized to the initial one in logarithm and in mass fraction as a function of the luminosity for various models.   
   The red full dots are for observed values, stars labeled 2, 3 and 6 are taken from \citet{Origlia13}, $\alpha$ Ori from \citet{Lambert84} and IRS 7 from \citet{Carr00}.
   {\it Upper right panel:} Same as Upper left panel but for the N/O surface abundance ratio. 
   {\it Lower left panel:} Tracks during the red supergiant phase in the N/C versus N/O plane (surface values). The models are the same as in the upper left figure.
        {\it Lower right panel:} Surface $^{12}$C/$^{13}$C ratios for the same models as in the upper left figure. The dots are observed values for 6, $\alpha$ Sco \citep{Hinkle76}, 3, 2, alpha Ori and VY CMa \citep{Matsuura14}.}
      \label{abond}
\end{figure*}
% macro/RSG/time15nc in PAPIER

%amŽliorer figure 12/13 en se concentrant sur le bas de la figure, ajouter labels aux Žtoiles

Figure~\ref{abond} shows how various abundance ratios evolve during the red supergiant phase for different initial masses, rotations and RSG mass-loss rates. 
The non-rotating tracks enter into the RSG with N/C, N/O and $^{12}$C/$^{13}$C ratios which are equal to the initial values (that means
that the tracks begin at the ordinate 0 in the two upper panels, at the coordinate (0,0) in the lower left panel and at the ordinate around 90, well outside the range of values shown in the lower right panel). Then the tracks
go up (increases of nitrogen and decreases of carbon and oxygen), except in the case of the lower right panel where the tracks go down when time increases ($^{12}$C/$^{13}$C decreases). 

Let us first focus on the case of the non-rotating 20 M$_\odot$ with standard mass-loss. We see that the RSG tracks cover a very large range of surface abundances, while the luminosity does not vary much.
The change of the surface abundances comes from the deepening of the outer convective zone that dredges-up some nuclearly processed material at the surface. For the N/C and N/O ratios, the changes are gradual and
the different ratios are relatively well distributed along the time sequence.  For the $^{12}$C/$^{13}$C ratio, the decrease is quite rapid and most of the time the model will show values below about 20. This is due to the fact that
the $^{12}$C/$^{13}$C ratio is significantly decreased on a greater extent of the mass of the star than the N/C or the N/O ratio.

 When the mass-loss rate is increased during the RSG phase, no significant differences appear for the 20 M$_\odot$ model. However,
due to the reduction  of the RSG lifetime when the mass-loss rates are larger, the ranges of surface abundance values are slightly more restricted. The same occurs for the luminosity.

When, for the 20M$_\odot$ model, rotational mixing is accounted for (continuous lines), we notice two significant differences with respect to the non-rotating cases. First, at the entrance of the RSG phase, the N/C, N/O and
$^{12}$C/$^{13}$C ratios are already reflecting the presence of nuclearly processed material at the surface. Second, the ranges of surface abundance ratios covered during the RSG phase are much smaller.
This comes from the fact that the outer convective zone in the rotating model does not extend as deep as in the non-rotating one, because in rotating models, helium cores are larger than in the non-rotating ones. As a 
consequence, the H-burning shell, which acts as a barrier for the deepening of the convective envelope, is at a larger  lagrangian mass coordinate in the rotating models.

Similar qualitative behaviors as for the 20 M$_\odot$ model are obtained for the 25 M$_\odot$ models. In the case of the 15 M$_\odot$ model, we just note that when the RSG mass-loss rate increases, then a larger range
of abundance ratios are obtained during the RSG phase for the N/C and N/O ratios. This comes from the fact that in the case of the 15 M$_\odot$ model, 
the mass lost during the RSG increases a lot when one passes from the standard to the enhanced RSG mass-losses (see also Fig.~\ref{Fig2TI}).
The 9 M$_\odot$ appears in  Fig.~\ref{abond} just through the standard mass-loss rate models (the part
of the track shown corresponds to the evolution back to the red after a blue loop.). The enhanced mass-loss rate models are at too low luminosities to show up on this plot. 

The lower left plot shows the track in the N/C versus N/O plane. We see that whatever the model considered the tracks are very similar. As discussed by \citet{Przybilla10} and \citet{Maeder14}, this results from the fact that
the tracks in this plane are more reflecting the CNO cycle itself than the details of the stellar models.

Recently surface abundances for three red supergiants belonging to the cluster RSGC2 have been obtained using the NIR spectrograph GIANO on Telescopio Nazionale Galileo (TNG) by \citet{Origlia13}.
The positions of these stars are indicated in Fig.~\ref{abond}.
RSGC2 is a young massive cluster (40 000 M$_\odot$) at a distance of about 3.5 kpc from the Galactic center \citep{Davies07}. This analysis finds that the [C/Fe] ratio is depleted by a factor between
two and three confirming the result by \citet{Davies09}. They also find that the $^{12}$C/$^{13}$C ratio is low (between 9 and 11). 
Values for the N/C and N/O values are also indicated for Betelgeuse and IRS 7 \citep{Lambert84, Carr00}. The $^{12}$C/$^{13}$C ratios are indicated for alpha Sco \citep{Hinkle76} and VY CMa \citep{Matsuura14}.
%A very low $^{12}$C/$^{13}$C ratio (5.6$\pm$1.8) was also obtained by Matsuura et al. (2013) using Herschel for the red supergiant VY CMa. This value is in the range
%of the $^{12}$C/$^{13}$C ratios of 3-14 obtained for four other red supergiants by Milam et al. (2009) but much lower than the estimates by these same authors for VY CMa (25-46) or by
%Nercessian et al. (1989) (36$\pm$9).

For the N/C and N/O ratios, with the exception of IRS 7, the ratios may be reproduced by the non-rotating models (whatever the mass-loss rate during the RSG stage) or by models
with an initial rotation smaller than the one corresponding to $\upsilon_{\rm ini}/\upsilon_{\rm crit}$ 0.4. The case of IRS 7, a star belonging to the galactic centre, does appear difficult to be explained by the present models.
It is also quite off from the predicted relations in the N/C versus N/O plane. The initial abundance ratios in the galactic centre may be different from solar ones, this might explain part of these differences.

We see that the observed  $^{12}$C/$^{13}$C ratios are all very small and appear slightly more compatible with models having undergone some mixing before entering the RSG stage.
This is in line with the conclusions by \citet{Davies09} and \citet{Origlia13}, and this conclusion is not changed considering modifications of the RSG mass-loss rates.
Actually we see that the surface abundances are much more sensitive to rotation than to the mass-loss rates during the RSG phase.  We can thus conclude that 
the constraints on the surface abundances of red supergiants cannot be used to discard any of the models presented here, even those computed with the most extreme mass-loss rates during the RSG phase.

% note sur C12C13
%For the 9 M$_\odot$ stellar models without rotation at solar metallicity, whatever the mass-loss rate during the RSG phase (1, 10 or 25 times) the range of $^{12}$C/$^{13}$C ratios obtained at the surface
%of the RSG is the same and vary from the initial value (at the base of the RSG branch) down to values around 20 at the tip of the RSG branch. 
%Is the decrease progressive? No it is a step function (dredge-up effect).
%What about now the 15 M$_\odot$? Qualitatively the same as for the 9 M$_\odot$, just that lower values of $^{12}$C/$^{13}$C are obtained, around 4. Idem 20 and 25.
%For the 25 it slightly changes the Teff at which a given 1213 is reached at the beginning of the RSG phase, but not very important.

%\subsubsection{Red supergiants and binarity}

\subsection{Surface and interior rotation of red supergiants}

During the RSG phase, the surface velocities at the equator are in general below 1 km s$^{-1}$. Surface velocities as low as 0.1 km s$^{-1}$ or even a few 0.01 km s$^{-1}$ can be reached. 
Models with enhanced mass-loss during the RSG phase present not significantly different surface velocities with respect to the standard mass-loss rate case.

The deprojected surface rotation velocity of Betelgeuse is about 15 km s$^{-1}$ \citep{Uitenbroek98}, so significantly above the values obtained by the present models, but still well below the typical critical velocity (between 40-60 km s$^{-1}$). 
 It is interesting to note that Betelgeuse is among the few red supergiants that are runaway stars. 
 Runaway stars move supersonically through the interstellar medium \citep{Blaauw61}.
 Such high speed may be acquired through few-body dynamical encounter \citep{Poveda67}, or binary-supernova explosions
\citep{Blaauw61, Stone91}. The supersonic movement produces an arc-like shape  bow shock that can be detected at many wavelengths
from infrared to X-ray wavebands. As written by \citet{Gvaramadze14}, most of bow-shock producing stars are either on the main-sequence or are blue supergiants,
while there are no Wolf-Rayet stars and only three among RSG: Betelgeuse \citep{Noriega97}, $\mu$ Cep \citep{Cox12} and IRC-10414 \citep{Gvaramadze14}.
Is the fast surface rotation observed for Betelgeuse related to the process which made it a runaway stars? We leave that question open. 

The ratio between the angular velocity of the core and of the surface is very large, being in the range of 10$^5$-10$^9$ for most models. Enhanced mass-loss rate models in general restrict
these ratios to values below 10$^7$.
Let us recall here that the present models
do not account for any additional transport mechanism in addition to those associated to shear turbulence and meridional currents. 
Asteroseismological analysis of red giants indicates that some additional transport mechanism is at work in stars with masses around 1 M$_\odot$  \citep{Beck2012, Eggen2012, Marques2013}. 
This additional mechanism produces stronger coupling between the core and envelope, reducing the ratio between the core and envelope rotation.
It would be very interesting to obtain for red supergiants similar constraints.

\subsection{Type II supernovae with a red supergiant progenitor}

%To have access to the positions of red supergiant progenitors of core collapse supernovae in the HRD is interesting to answer two
%questions: 1) Which kind of supernovae is produced by RSG progenitors? 2) What is the mass domain of stars that have their end point of their evolution in
%the RSG phase? The answer of the first question is fairly well known. 

%Thus 
%theory as well as observation indicate that most if not all RSG produce type IIP supernovae.
%The answer of the second question is less well known and is the main topic of the present section.

\begin{figure*}
%: fig TcrcZ014_Mcc_rot.eps
\centering
\includegraphics[width=.47\textwidth, angle =0]{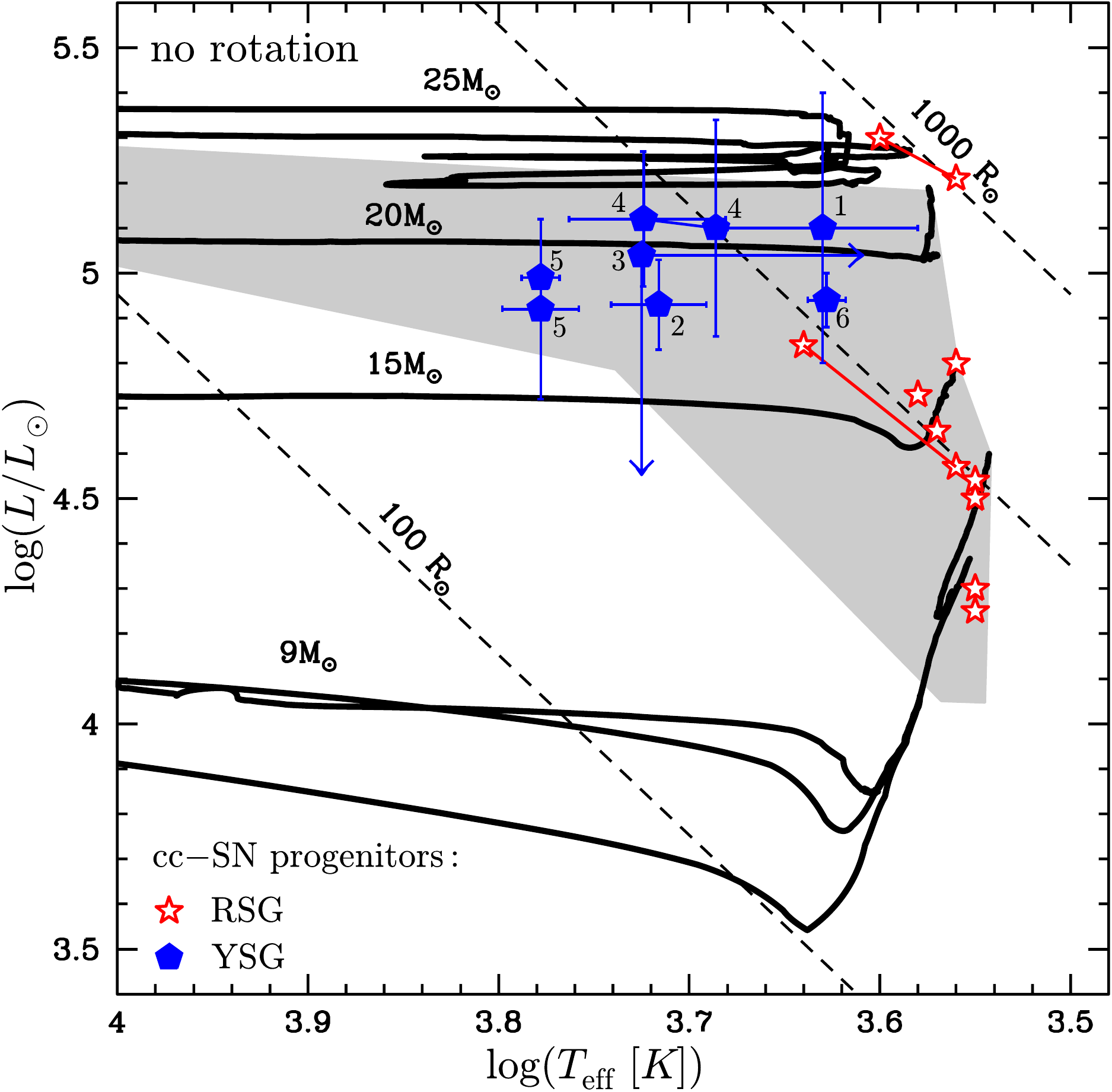}\includegraphics[width=.47\textwidth, angle =0]{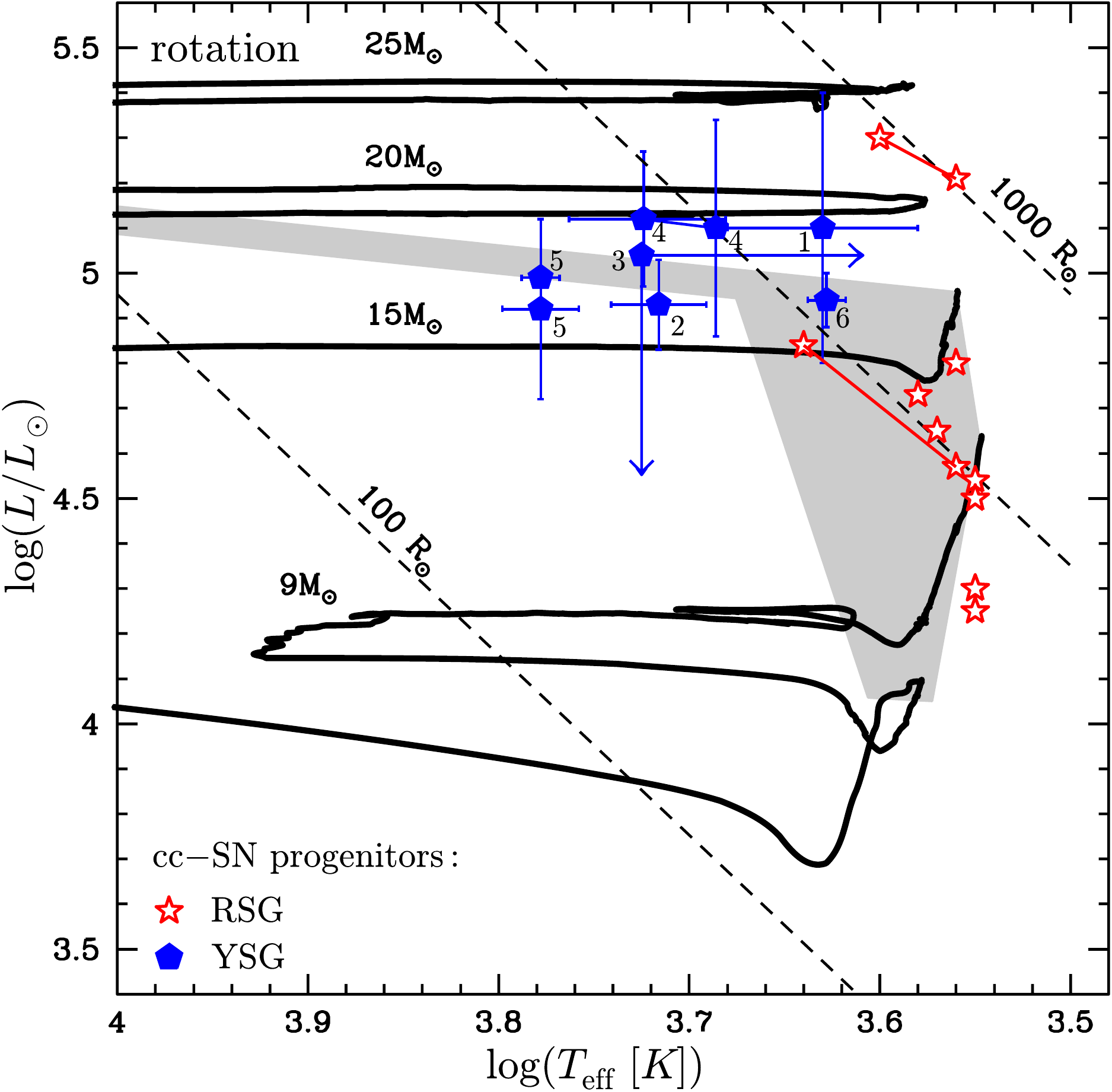}
   \caption{{\it Left panel:} Evolutionary tracks in the HR diagram for the non-rotating models and computed with the standard mass-loss rates 
   with superposed the positions of progenitors of core collapse supernovae. Empty stars are for those which are red supergiants (see Table~\ref{tab2}),
   filled pentagons are for those which have a yellow supergiant as progenitors (see Table~\ref{TableIIYSG}).
   A continuous segment links positions of the same SN progenitors obtained by various authors.
 Progenitors of supernovae predicted by the non-rotating models computed with the various
   RSG mass loss rates are found in the shaded area. Progenitors obtained with standard RSG mass-loss rates occupies the upper part as well as the right part of the shaded region, while
   the progenitors obtained from enhanced RSG mass-loss rate models are on the lower-left region of the shaded area.
   Lines of constant radius are indicated, the line intermediate between the 100 and 1000 R$_\odot$ corresponds to a radius of 500 R$_\odot$. {\it Right panel:} Same as left panel with evolutionary tracks in the HR diagram for the rotating models and computed with the standard mass-loss rates.
   Progenitors of supernovae predicted by the rotating models computed with the various
   RSG mass loss rates are found in the shaded area.}
      \label{snrsg}
\end{figure*}

%\begin{figure}[t]
%: fig TcrcZ014_Mcc_rot.eps
%\centering
%\includegraphics[width=.51\textwidth, angle =0]{figHRrcc.eps}
%   \caption{Same as Fig.~\ref{snrsg} with evolutionary tracks in the HR diagram for the rotating models and computed with the standard mass-loss rates.
%   Progenitors of supernovae predicted by the rotating models computed with the various
%   RSG mass loss rates are found in the shaded blue area.}
%      \label{snyel}
%\end{figure}
\begin{figure*}
%: fig TcrcZ014_Mcc_rot.eps
\centering
\includegraphics[width=.81\textwidth, angle =0]{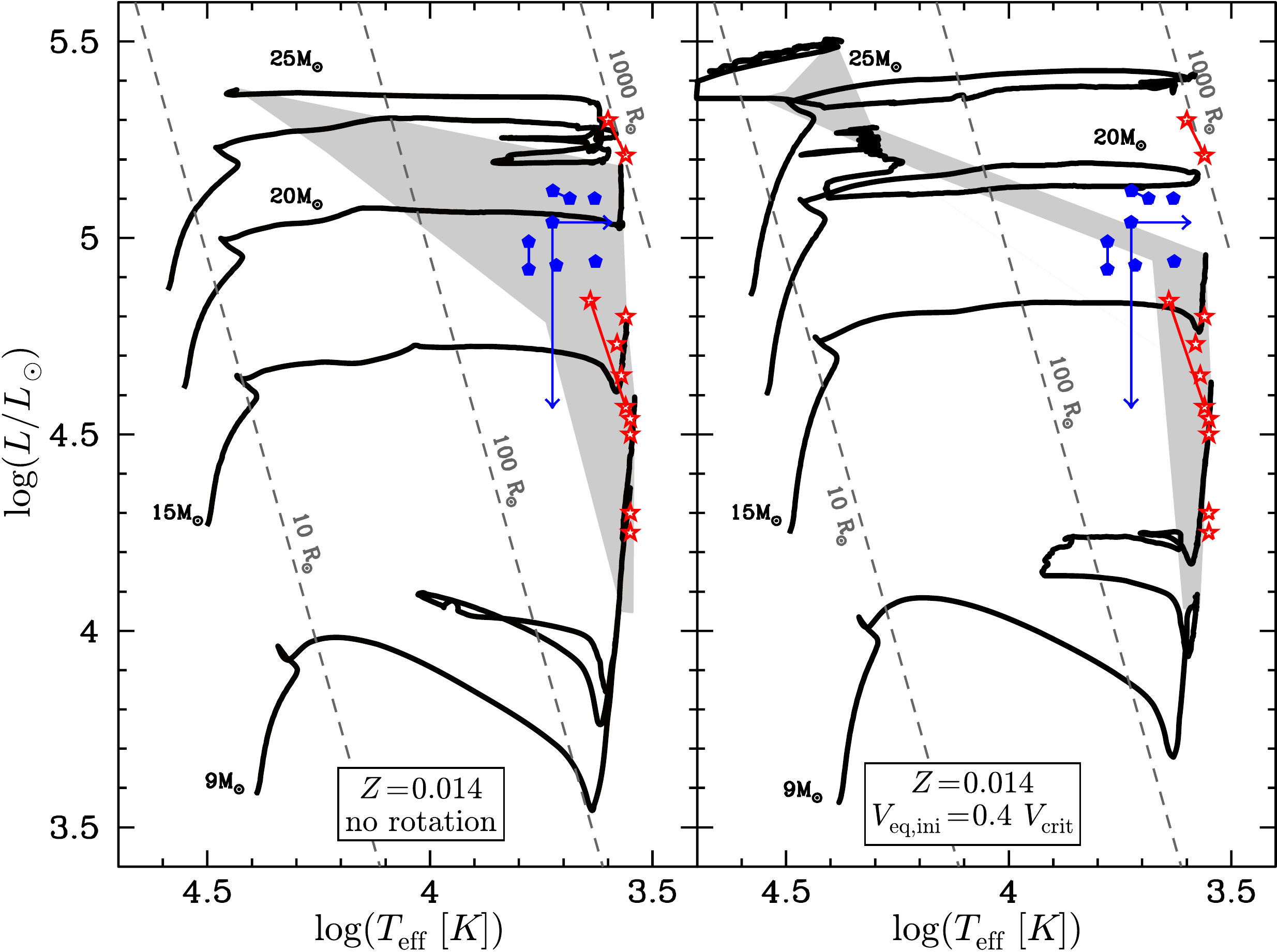}
   \caption{{\it Left panel:} Evolutionary tracks in the HR diagram for the non-rotating models and computed with the standard mass-loss rates with superposed observed positions of progenitors of core collapse supernovae.
   The observed data are the same as in Figs.~\ref{snrsg}. For purpose of clarity, the error bars are not shown. 
   Progenitors of supernovae predicted by the non-rotating models computed with the various
   RSG mass loss rates are found in the shaded area.
  {\it Right panel:} Same as the left panel for rotating stellar models. The shaded area shows the region where the progenitors of supernovae are found according to the rotating models with various RSG mass loss rate prescriptions.}
      \label{Fig4HR}
\end{figure*}

\begin{table}
\caption{Properties of RSGs identified as SN progenitors. All these supernovae are of type II-P.}
\label{tab2}
\centering
\scriptsize{
\begin{tabular}{ccccc}
\hline\hline
    SN  & [O/H] & Log L/L$_\odot$  & Log (T$_{\rm eff}/K)$ & Reference \\
\hline
2003gd  & 8.4                    &  4.3$\pm$0.3     & 3.55                   & \citet{Smartt2009} \\ 
2004A   & 8.3                     &  4.5$\pm$0.25   & 3.55                   & \citet{Smartt2009}\\ 
2004et  & 8.3-8.9               &  4.8                    & 3.56                    & Fraser et al. (in prep.)\\ 
2005cs & 8.7                      &  4.25$\pm$0.25 & 3.55                    & \citet{Smartt2009} \\
2008bk & 8.4-8.7                &  4.84$\pm$0.12 & 3.64                     & \citet{Maund2014}\\
              & 8.5                     &  4.57$\pm$0.06 & 3.56$\pm$0.006  & \citet{vandyk2012}\\
2009md & 8.96$\pm$0.04  & 4.54 $\pm$0.19 & 3.55$\pm$0.010    & \citet{Fraser2011}\\
2012A    & 8.12$\pm$0.08  &  4.73$\pm$0.14 &  3.58$\pm$0.050   & \citet{Tomasella2013}\\
2012aw &  8.6$\pm$0.2     &   5.3$\pm$0.3     & 3.60$\pm$0.050    & \citet{Fraser2012}\\
             &  8.7                     &  5.21$\pm$0.03 & 3.56$\pm$0.025   & \citet{VANDYK12aw}\\
2013ej  &  -                         & 4.65$\pm$0.20  & 3.57$\pm$0.035   & \citet{Fraser2014}\\              
\hline
\end{tabular}
}
%\label{RSGSN}
\end{table}

\begin{table}
\caption{Durations of various post-MS  stages in Myr.}
\label{TabBR}
\centering
\scriptsize{
\begin{tabular}{ccccccccc }
\hline\hline
    M$_{\rm ini}$ &  t$_{B1}$ & t$_{Y1}$ & t$_{RSG}$ &  t$_{Y2}$ & t$_{B2}$ &   t$_{WR}$ & t$_{pRSG}$ & t$_{pMS}$ \\

 \hline
 & & & & & & & &    \\ 
 \multicolumn{9}{c}{$\upsilon_{\rm ini}$=0.} \\
 & & & & & & & &   \\
   \multicolumn{9}{c}{1$\times \dot{M}_{\rm stan.}$ } \\      
   15    &   0.210 &  0.150 & 1.008 & 0       &  0      &     0      &   0         & 1.580 \\
   20   &   0.271 &  0.092 & 0.549 & 0       &  0      &     0      &   0         & 0.912 \\
   25   &   0.100 &  0.088 & 0.296 & 0.233 &  0     &     0.001 &   0.234  & 0.718 \\
 & & & & & & & &   \\
  \multicolumn{9}{c}{10$\times \dot{M}_{\rm stan.}$ } \\    
   15   &  0.210 &  0.150 & 0.703 & 0.393 &  0        &     0      &   0.393  & 1.456 \\ 
   20   &  0.271 &  0.092 & 0.187 & 0.357 &  0       &     0      &   0.357  & 0.907 \\ 
   25   &   0.100 &  0.087 & 0.052 & 0.093 &  0.361 &     0     &   0.454  & 0.694 \\    
 & & & & & & & &   \\
  \multicolumn{9}{c}{25$\times \dot{M}_{\rm stan.}$ } \\ 
   15   &  0.210 &  0.150 & 0.293 & 0.502 &  0.286 &     0      &   0.788 & 1.440 \\  
   20   &  0.271 &  0.092 & 0.074 & 0.194 &  0.269 &     0      &   0.463 & 0.900 \\ 
   25   &  0.100 &  0.103 & 0.012 & 0.248 &  0.238 &     0      &   0.486 & 0.701 \\  
   & & & & & & & &    \\
 \multicolumn{9}{c}{$\upsilon_{\rm ini}$=0.4$\upsilon_{\rm crit}$} \\
 & & & & & & & &    \\ 
    \multicolumn{9}{c}{1$\times \dot{M}_{\rm stan.}$ } \\           
   15    &  0.060 &  0.110 & 1.193 & 0       &  0        &     0  &   0         & 1.363 \\
   20   &  0.051 &  0.157 & 0.345 & 0.056 &  0.271 &   0    &   0.326  & 0.880 \\
   25   &  0.014 &  0.003 & 0.130 & 0.080 &  0.415 &   0    &   0.495  & 0.641 \\
 & & & & & & & &  \\ 
   \multicolumn{9}{c}{10$\times \dot{M}_{\rm stan.}$ } \\       
   15   &  0.060 &  0.108 & 0.430 & 0.729 &  0.117 & 0        &   0.847  & 1.445 \\ 
   20   &  0.051 &  0.157 & 0.111 & 0.037 &  0.516 &     0      &   0.553  & 0.872 \\ 
   25   &  0.014 &  0.004 & 0.033 & 0.103 &  0.475 &     0     &   0.578  & 0.632 \\    
 & & & & & & & &   \\ 
  \multicolumn{9}{c}{25$\times \dot{M}_{\rm stan.}$ } \\ 
   15   &  0.060 &  0.107 & 0.193 & 0.732 &  0.184 &     0      &   0.916 & 1.276 \\  
   20   &  0.051 &  0.158 & 0.045 & 0.036 &  0.583 &     0      &   0.619 & 0.873 \\ 
   25   &  0.014 &  0.004 & 0.012 & 0.098 &  0.506 &     0      &   0.604 & 0.634 \\  
 & & & & & & & &   \\      
 \hline
\end{tabular}
%\tablefoot{The initial masses and the elements quantity are given in M$_{\sun}$.\\
%}
}
\end{table} 

In Table~\ref{tab2}, we list some properties of the RSGs identified as SN II-P progenitors according to the most recent determinations in the literature.
Only those progenitors of type II-P supernovae for which the luminosity is available are indicated, letting aside those for which only an upper limit is given.
Note that when no effective temperature is given, an arbitrary value of 3.55 is attributed to the progenitor, based on the fact that
according to the references quoted, the progenitor was likely a red supergiant.
The metallicities given are very indicative. 

%\sout{To build this table, we started from the list given by \citet{Smartt2009MNRAS} \citep[see also the discussion in][]{Smartt2009},\sout{ who mention 6 supernovae (1999ev, 2003gd, 2004A, 2004et, 2005cs and 2008bk) which have a visible
%potential progenitor in pre-explosion archive images. 
%According to \citet{Maund2014MNRAS}, the progenitor of 1999ev is not the one identified by  \citet{Smartt2009MNRAS},
%thus we did not list it in Table~\ref{tab2}. 
%The properties of two progenitors studied by  \citet{Smartt2009MNRAS} have been reanalyzed recently. Fraser et al. (in preparation\footnote{the values shown in the table correspond to values reported by \citet{Jerkstrand2012}, who refers to a still unpublished result by Fraser et al.})
%obtain for 2004et a higher luminosity 
%than the one quoted by \citet{Smartt2009MNRAS} by about 0.2 dex.
%The progenitor of 2008bk was reanalyzed by two teams, \citet{Maund2014} and \citet{vandyk2012}. The effective temperature obtained by \citet{Maund2014}
%would classify this progenitor more as a yellow than a red supergiant, while the effective temperature obtained by \citet{vandyk2012} would keep it among the
%red supergiant progenitors. 
%In addition to these updates, four new progenitors have been detected since the work by \citet{Smartt2009MNRAS}.
%The positions of all these progenitors are shown in Figs.~\ref{snrsg} and \ref{snyel} by stars. 
%When more than one value are given in the literature, we show the two values connected by a line.}

Looking at the positions in Figs.~\ref{snrsg} of these nine progenitors, we can make the following comments:
\begin{itemize}
\item Most of the progenitors have Log $L/L_\odot$ between 4.2 and 4.8, which means  well within the luminosity range of stellar models ending
their lifetimes as red supergiants when standard mass-loss rates are used. This holds for both rotating and non-rotating models. Models with the mass-loss rate increased by a factor of 10 can barely match the observed RSG at the pre-SN stage, meaning that we cannot rule out that models with modest mass-loss enhancements $\dot{M}$ (2--4) would be able to fit the observations.
On the other hand, these progenitors cannot be fitted by models with the mass-loss rate enhanced by a factor of 25, which predict too low luminosities or too high effective temperatures in this luminosity range. 
\item  As discussed in \citet{Groh2013}, the initial mass of observed progenitors that are RSGs depend on rotation, with non-rotating models yielding a larger initial mass than rotating models. We refer to their Table 6 for determinations of the initial mass of SN II-P progenitors based on rotating and non-rotating models. 
% \sout{ The present grid is too coarse for associating very precise initial mass to each of the
%progenitors, but, as an example, the initial mass of the progenitor for 2004et would be a 15 M$_\odot$ in case non-rotating models are considered
%and 11-12 M$_\odot$ in case rotating models are considered. The non-rotating model is favored however.}
\item For SN 2004et, the nebular-phase spectral modeling made by \citet{Jerkstrand2012} constrains the progenitor mass to M$_{\rm ZAMS}$ = 15 M$_\odot$, with a pre-SN oxygen mass of 0.8 M$_\odot$. This oxygen mass is quite consistent with our non-rotating 15 M$_\odot$ model with standard mass-loss rate during the RSG phase, which predicts that an oxygen mass of 0.8 M$_\odot$ would be ejected assuming a remnant mass of about 1 M$_\odot$.
\item The case of 2012aw (the most luminous progenitor) could be explained by our non-rotating standard mass-loss rate model for a 20 M$_\odot$ star or from a model with a slightly higher initial mass
(although less massive than 25 M$_\odot$). It appears more difficult to fit that progenitor from rotating models with standard mass-losses.
Such a highly luminous RSG progenitor cannot be reproduced by enhanced mass-loss rate models, with or without rotation.
More generally, the upper luminosity of stars ending their evolution as a red supergiant 
decreases when the initial rotation 
and/of the mass-loss rate during the red supergiant stage increases. This can well be seen in Fig.~\ref{Fig4HR}.
\item For the 9~M$_\odot$ models, an enhancement in the mass-loss rate at the RSG phase decreases the luminosity at the pre-SN stage. This may have interesting consequences for the nature of low-luminosity SN II-P progenitors and the minimum initial mass of stars that produce core-collapse SNe.
\end{itemize}
From the points above, we conclude that the positions in the HRD of the present RSG supernova progenitors are 
best described by the standard mass-loss models (see Sect. 8.1 of \citealt{Groh2013}), although modest mass-loss rate enhancements (2--4) cannot be discarded.

%discussion vitesse de rotation de Betelgeuse, champ magnŽtique extŽrieur, buffer envelope, ajouter vitesse de surface table 1. rapport Vcoeur/Venvelope

\section{Evolution of post-RSG stars}

On Fig.~\ref{hrrsgwr}, we can see how the post red supergiant tracks change when different RSG mass-loss rates are applied. If we focus on the case of the 20 M$_\odot$ model, we can note
that increasing the RSG mass-loss extends the post RSG track towards bluer positions.
For the 15 M$_\odot$ model,  a qualitatively similar behavior is observed although less marked. In the case of the 25 M$_\odot$, on the other hand, we obtain that
the evolution back to the blue, which was already present in the standard mass-loss rate model, is slightly shortened in its extension in effective temperatures and does occur at lower luminosities when enhanced mass-loss rate models are used. {For all the cases discussed above (15 to 25~M$_\odot$), the star would explode as an LBV or yellow hypergiant \citep{gme13,Groh2013}.}
Finally, as already explained in Section 3, increasing the RSG mass-loss rate for the 9 M$_\odot$ keeps the star in the red part of the HRD. 
When rotation is accounted for, these features remain very similar.

Table~\ref{TabBR} shows the time spent by the present stellar models in the blue, yellow, red regions of the HR diagram, and as Wolf-Rayet stars. Following the recent results from \citet{Groh2014}, we consider the models as being WR stars when $\log (T_{\rm eff}/K)>4.36$ and the mass fraction of hydrogen at the surface is less than 0.30. We consider the model as being in the blue region of the HR diagram when $\log (T_{\rm eff}/K)> 3.90$ and when it is not a WR star. Our criteria for determining the blue region include blue supergiants, blue hypergiants, and LBVs.
We distinguish the time spent in the blue before ($t_{B1}$) and after the RSG phase ($t_{B2}$). Yellow stars are those
stars with $3.66<\log (T_{\rm eff}/K)< 3.90$, which encompasses yellow supergiants and yellow hypergiants. Also here, we note $t_{Y1}$, respectively $t_{Y2}$, the duration of the yellow
phase before and after the RSG phase.  Red supergiants are considered as those stars with $\log (T_{\rm eff}/K)< 3.66$ for the initial stellar masses shown in Table~\ref{TabBR}.

Globally, we see that the total time spent after the Main Sequence phase (see $t_{\rm pMS}$ in Table~\ref{TabBR}) is not much affected by a change of the RSG mass-loss rate.
As already mentioned before, the change of the RSG mass-loss rate has a deep impact on the duration of the post RSG phase (see $t_{\rm pRSG}$ in Table~\ref{TabBR}).
As an example, the 15 M$_\odot$ with standard mass-loss rate spend no time in post RSG phases, while the enhanced mass-loss rate model spends a fraction between 27 and 72\% of the whole
post MS period.

The fraction of the post MS phase that is spent in the yellow and blue regions depends on the initial mass, rotation and the RSG mass-loss rates.  
When the stellar model evolves back to the blue, then the duration of the post-RSG yellow or blue supergiant phase is in general larger
than the duration of the corresponding phases before the RSG stage. Therefore, when a blue ward evolution occurs, there is a greater chance that a given blue or yellow supergiant be a post RSG object than a pre RSG one. 
%As mentioned above, the post RSG phase is a significant fraction
%of the total post MS phase, it means that a change of the RSG mass-loss rate has a significant impact on the populations of stars in the yellow and blue parts of the HR diagram.
%Typically, for the non-rotating (rotating) 20 M$_\odot$ model, the ratio $(t_{B1}+ t_{B2})/t_{\rm RSG}$ passes from a value equal to 0.5 (0.9) for the standard model to 1.4 (5.1) and 7.3 (14.2) for 
%respectively the 10 and 25 $\times$ the RSG mass-loss rate models. 
%5The ratios $t_{B2}$/$t_{B1}$ are in general equal or larger than 2.0 when the values are not zero. 
%This means that, when an extended blueward evolution is obtained after a RSG phase, the fraction of blue stars in a post-RSG stage over the total number of blue stars would be 50\% or higher. 
 
The enhanced mass-loss models never enter into the WR stage\footnote{This of course is dependent on the way we define a WR star but the feature that would remain is the fact that
whatever the type of the star, the surface composition of the enhanced RSG mass-loss models correspond in general to less CNO processed material than the surface composition of the standard models.}! This might be surprising at first sight since one would have expected that stronger mass-losses during the RSG phase would
favor the formation of WR stars. Actually, as noted above, when the mass-loss rate is increased, the duration of the RSG phase is shortened while the total mass lost remains more or less constant. As a consequence the
star with enhanced mass-loss rate will expose at the surface more or less the same interior layer but at an earlier stage of its evolution. This explains why the surface composition will reflect a less advanced evolutionary phase.
%However, we find that increasing the RSG mass-loss rate does not change the total mass lost during the RSG phase (see Fig.~\ref{Fig2TI}). What is changed on the other hand is the duration of the RSG phase. Therefore,
%when the star evolves away from the RSG phase, 
%the total mass of the star will be more or less the same for the range of RSG mass-loss rates studied here, and its chemical structure will be a little different. The models with enhanced mass-loss present a slightly less evolved chemical structure. As already indicated above, the stellar winds during the post RSG phase remain modest in our models and will not much change the situation, thus one expects surface abundances reflecting a slightly less evolved state in enhanced mass-loss rate models than in standard ones. This explains why for enhanced mass-loss rate models, we have more difficulties
%to enter the WR phase. 
So, increasing the mass-loss rate during the RSG phase, would actually decrease the number of WR stars formed through the single star channel!
A caveat concerns the occurrence of eruptions during the post-RSG phase, when the star becomes a yellow hypergiant or an LBV \citep{Groh2013}, which could remove additional mass from the star and favor the formation of WRs. Therefore, it seems that the key for forming WR stars at low-luminosity from single stars is the post-RSG mass-loss.

\subsection{Effective temperatures and characteristics of SN progenitors}

\begin{table*}
\caption{Characteristics of the models at the end of the core He- or C-burning phase (models at the end of the C-burning phase are indicated by an asterisk).}
\label{Tab2}
\scriptsize{
\centering
\begin{tabular}{r r r r r r r r r r r r r r r r}
\hline\hline
    M$_{\rm ini}$ & $\dot{M}$                 & R$_{\rm fin}$ & M$_{\rm fin}$ & $\Delta$M$_{\rm H}$ & M$_\alpha$ & M$_{\rm CO}$ & M$_{\rm rem}$ & M$_{\rm grav}$ & m$_{\rm H}$  & m$_{\rm He}$  &  m$_{\rm C}$ & m$_{\rm O}$ & L   & $\Omega$ (NS) & P (NS)  \\
    M$_\odot$     &  M$_\odot$ y$^{-1}$& R$_\odot$   & M$_\odot$   & M$_\odot$       & M$_\odot$   & M$_\odot$       &  M$_\odot$       &M$_\odot$          & M$_\odot$     & M$_\odot$        & M$_\odot$     & M$_\odot$    & $10^{49}$                 & 10$^4$              & $10^{-4}$  \\
                          &                                 &           &    &   &            &                         &                          &                           &                        &                         &                       &                      & cm$^2$ g/sec           &  sec$^{-1}$         & sec \\
 \hline
 & & & & & & & & & & & &  & & & \\ 
 \multicolumn{16}{c}{$\upsilon_{\rm ini}$=0.} \\
 & & & & & & & & & & & &  & & & \\      
    9    & 1$\times \dot{M}_{\rm stan.}$   & 548 & 8.765 & 7.555 &  1.21     & 1.20 & 1.12   & 1.05   & 4.58 & 2.87 & 0.05 & 0.07 & --      & -- & --  \\
   15   & 1$\times \dot{M}_{\rm stan.}$   &  637 & 13.174 & 8.914 & 4.26     & 2.24 & 1.46   & 1.33   & 5.76 & 4.94 & 0.37 & 0.52 & --      & -- & --  \\
   20   & 1$\times \dot{M}_{\rm stan.}$   & 932 & 8.635 & 2.425 &  6.21     & 4.00 & 1.91   & 1.68   & 1.15 & 3.31 & 0.49 & 1.16 & --      & -- & --  \\
   25   & 1$\times \dot{M}_{\rm stan.}$   & 22   & 8.289 & 0.169 & 8.12     & 5.95 & 2.41   & 2.03   & 0.03 & 2.20 & 1.01 & 2.28 & --      & -- & --  \\    
 & & & & & & & & & & & &  & & &   \\    
    9    & 10$\times \dot{M}_{\rm stan.}$   & 258 & 6.688 & 4.558 & 2.13     & 0.88 & 0.88   & 0.84   & 3.19 & 2.43 & 0.10 & 0.05 & --      & -- & --  \\
   15   & 10$\times \dot{M}_{\rm stan.}$   & 491 & 4.856 & 0.616 & 4.24     & 2.24 & 1.46   & 1.33   & 0.28 & 2.01 & 0.52 & 0.55 & --      & -- & --  \\
   20   & 10$\times \dot{M}_{\rm stan.}$   & 620 & 6.641 & 0.541 & 6.10     & 3.81 & 1.86   & 1.65   & 0.27 & 2.38 & 0.75 & 1.30 & --      & -- & --  \\
   25   & 10$\times \dot{M}_{\rm stan.}$   & 47 & 8.265 & 0.365 & 7.90     & 5.54 & 2.30   & 1.96   & 0.16 & 2.44 & 1.06 & 2.17 & --      & -- & --  \\    
 & & & & & & & & & & & &  & & &  \\ 
     9    & 25$\times \dot{M}_{\rm stan.}$  &  287 & 3.471 & 1.361 & 2.11     & 0.87 & 0.87   & 0.83   & 0.93 & 1.55 & 0.07 & 0.02 & --      & -- & --  \\
   15   & 25$\times \dot{M}_{\rm stan.}$   &  276 & 4.601 & 0.401 & 4.20     & 2.21 & 1.45   & 1.32   & 0.21 & 1.98 & 0.44 & 0.53 & --      & -- & --  \\
   20   & 25$\times \dot{M}_{\rm stan.}$   &  148 & 6.501 & 0.461 & 6.04     & 3.75 & 1.85   & 1.64   & 0.22 & 2.32 & 0.76 & 1.28 & --      & -- & --  \\
   25   & 25$\times \dot{M}_{\rm stan.}$   &  49 & 8.284 & 0.264 & 8.02     & 5.66 & 2.33   & 1.98   & 0.14 & 2.41 & 1.07 & 2.24 & --      & -- & --  \\    
& & & & & & & & & & & &   & & & \\
 \multicolumn{16}{c}{$\upsilon_{\rm ini}$=0.4$\upsilon_{\rm crit}$} \\
 & & & & & & & & & & & &   & & & \\      
    9$^{*}$  & 1$\times \dot{M}_{\rm stan.}$   &  555 & 8.517 & 5.437 & 3.08     & 1.64 & 1.30   & 1.20   & 3.53 & 3.03 & 0.18 & 0.34 & 6.24    & 5.18 & 1.21  \\
   15   & 1$\times \dot{M}_{\rm stan.}$   &  766 & 11.516 & 6.506 & 5.01     & 2.78 & 1.60   & 1.44   & 3.72 & 4.62 & 0.48 & 0.97 & 9.50    & 6.56 & 0.96 \\
   20$^{*}$ & 1$\times \dot{M}_{\rm stan.}$   &  35 & 7.178 & 0.008 & 7.17     & 4.73 & 2.10   & 1.82   & 0.02 & 1.61 & 0.87 & 2.10 & 11.87  & 6.52 & 0.96  \\
   25$^{*}$   & 1$\times \dot{M}_{\rm stan.}$   &  31 & 9.690 & 0.000 &9.69     & 7.09 & 2.69   & 2.22   & 0.00 & 1.59 & 1.61 & 3.60 & 17.77  & 7.98 & 0.79  \\    
 & & & & & & & & & & & &   & & & \\    
    9    & 10$\times \dot{M}_{\rm stan.}$   &  254 & 6.111 & 3.811 & 2.30     & 0.96 & 0.96   & 0.91   & 2.48 & 2.44 & 0.11 & 0.07 & 4.43     & 4.84 & 1.30  \\
   15   & 10$\times \dot{M}_{\rm stan.}$   &  442 & 5.306 & 0.386 & 4.92     & 2.79 & 1.60   & 1.44   & 0.15 & 1.89 & 0.55 & 1.05 & 2.96     & 2.05 & 3.07  \\
   20   & 10$\times \dot{M}_{\rm stan.}$   &  47 & 7.297 & 0.337 & 6.96     & 4.86 & 2.06   & 1.79   & 0.15 & 2.21 & 0.89 & 1.93 & 4.89     & 2.50 & 2.51  \\
   25   & 10$\times \dot{M}_{\rm stan.}$   &  16 & 9.715 & 0.245 & 9.47    & 7.02 & 2.67   & 2.21   & 0.09 & 2.42 & 1. 35 & 3.05 & 0.38      & 0.17 & 36.45 \\    
 & & & & & & & & & & & &   & & & \\ 
     9    & 25$\times \dot{M}_{\rm stan.}$  &  219 & 2.595 & 0.325 & 2.27     & 0.94 & 0.94   & 0.89   & 0.21 & 1.32 & 0.07 & 0.04 & 4.34   & 4.85 & 1.29  \\
   15   & 25$\times \dot{M}_{\rm stan.}$   &  557 & 5.309 & 0.609 & 4.70     & 2.71 & 1.58   & 1.43   & 0.24 & 2.03 & 0.61 & 0.78 & 7.72  & 5.39 & 1.16  \\
   20   & 25$\times \dot{M}_{\rm stan.}$   &  49 & 7.189 & 0.299 & 6.89     & 4.52 & 2.04   & 1.78   & 0.14 & 2.17 & 0.89 & 1.88 & 4.79  & 2.68 & 2.34 \\
   25   & 25$\times \dot{M}_{\rm stan.}$   &  12 & 9.620 & 0.230 & 9.39     & 6.88 & 2.64   & 2.18   & 0.86 & 2.32 & 1.37 & 3.03 & 0.38  & 0.17 & 36.37\\    
 & & & & & & & & & & & &  & & & \\      
 \hline
\end{tabular}
%\tablefoot{The initial masses and the elements quantity are given in M$_{\sun}$.\\
%}
}
\end{table*}

\begin{figure}
%: fig TcrcZ014_Mcc_rot.eps
\centering
\includegraphics[width=.49\textwidth]{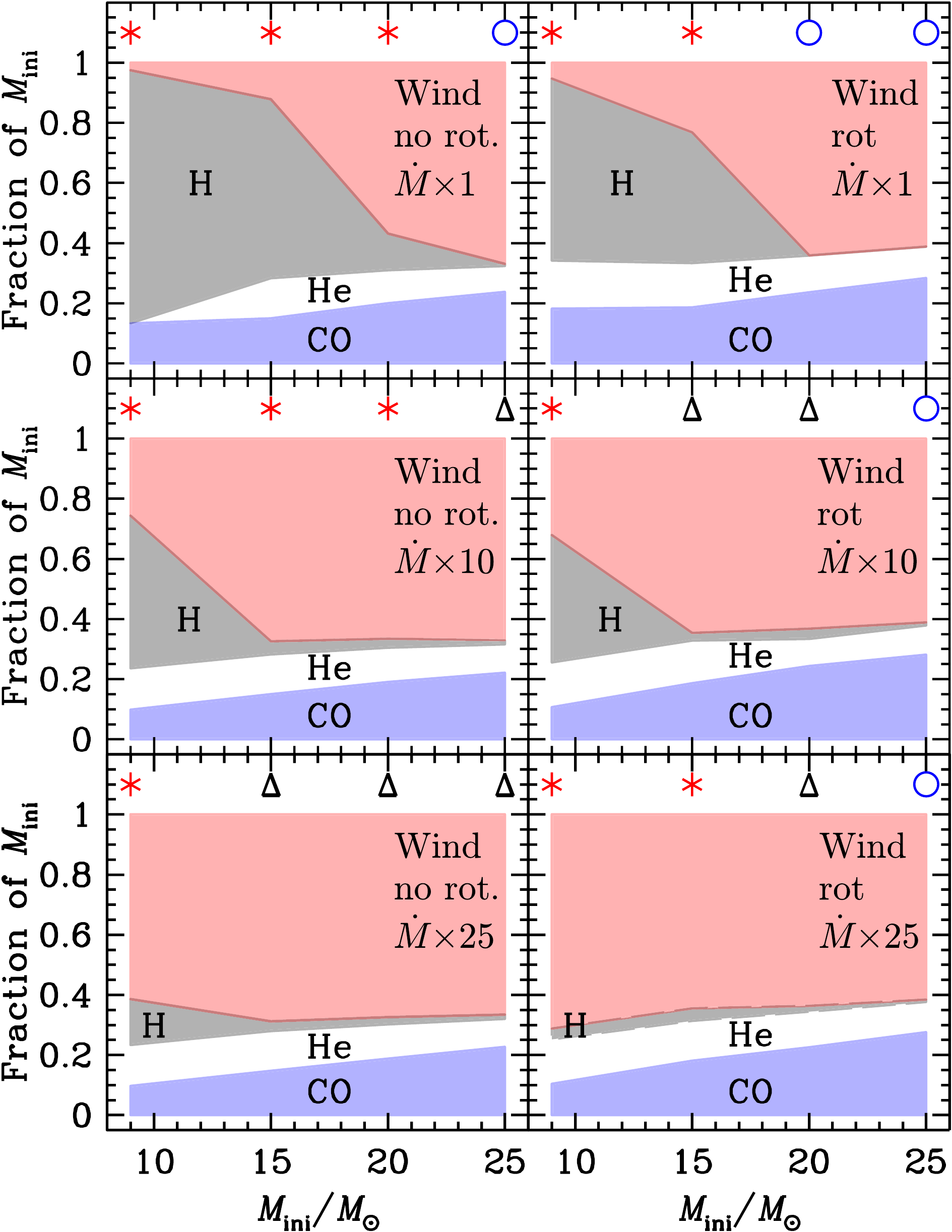}
   \caption{Fraction of the initial mass of the star ejected by stellar winds during the whole stellar lifetime (red shaded area). The mass in the H-rich envelope, He-rich region and CO core in the pre-supernova model as a function of the initial mass are indicated by the black, white and blue regions. Cases for various mass-loss rates during the RSG phase without and with rotation are shown. The symbols in the upper part of each panel indicate the range of effective temperatures of the model at the pre-supernova stage: log T$_{\rm eff}$ inferior to 3.65 are indicated by (red) stars, log T$_{\rm eff}$ between 3.65 and 4.30 by (black) triangles and log T$_{\rm eff}$ greater than 4.3 by (blue) circles.}
      \label{COHEf}
\end{figure}

In Fig~\ref{Fig4HR}, we show  the
regions in the HR diagrams (see the shaded areas), where the models computed with different RSG mass-loss rates predict the positions of the SNe progenitors\footnote{The models were computed until the end of the core C- or He-burning phase. The positions of the models at the end of the He-burning phase can still evolve during the core carbon burning phase, but
in absence of very strong mass-loss outburst, the displacement in the HR diagram remains modest and we shall consider the positions of these models as a good proxy for the
pre-supernova positions.}.  When the RSG mass-loss rate is increased, as also discussed by \citet{Georgy2012}, the end point of the evolution is shifted to the blue for the 15 and 20 M$_\odot$ models.
%We see that for the 15 M$_\odot$ model, the end point is only slightly blue shifted even for very large RSG mass-loss rates. This comes from the limitation in the total mass that can be lost during the RSG phase
%due to the negative feedback of an increased mass-loss rate on the duration of the RSG phase. Actually, as can be see from Fig.~\ref{Fig4HR}, the final masses are nearly the same for a RSG mass-loss rate increase by a factor of
%10 or 25. The same is true for the 20 M$_\odot$ although here the shift to the blue is much larger. {\bf CHECK THIS, I HAVE 9 Msun lost during the RSG for the same model!!! }Actually, for this mass, the mass lost during the RSG phase is modest (only 0.3 M$_\odot$) but it has
%a very strong impact on the position of the pre-supernova model in the HR diagram. So from these models, on would conclude that the mass domain between about 15 and 20 M$_\odot$ is likely the most sensitive one to a change
%of RSG mass-loss rate. 
%When considering the impact on evolutionary tracks, the mass domain between about 15 and 20 M$_\odot$ is the most sensitive one to a change
%of RSG mass-loss rate.

\begin{figure}
%: fig TcrcZ014_Mcc_rot.eps
\centering
\includegraphics[width=.49\textwidth, angle =0]{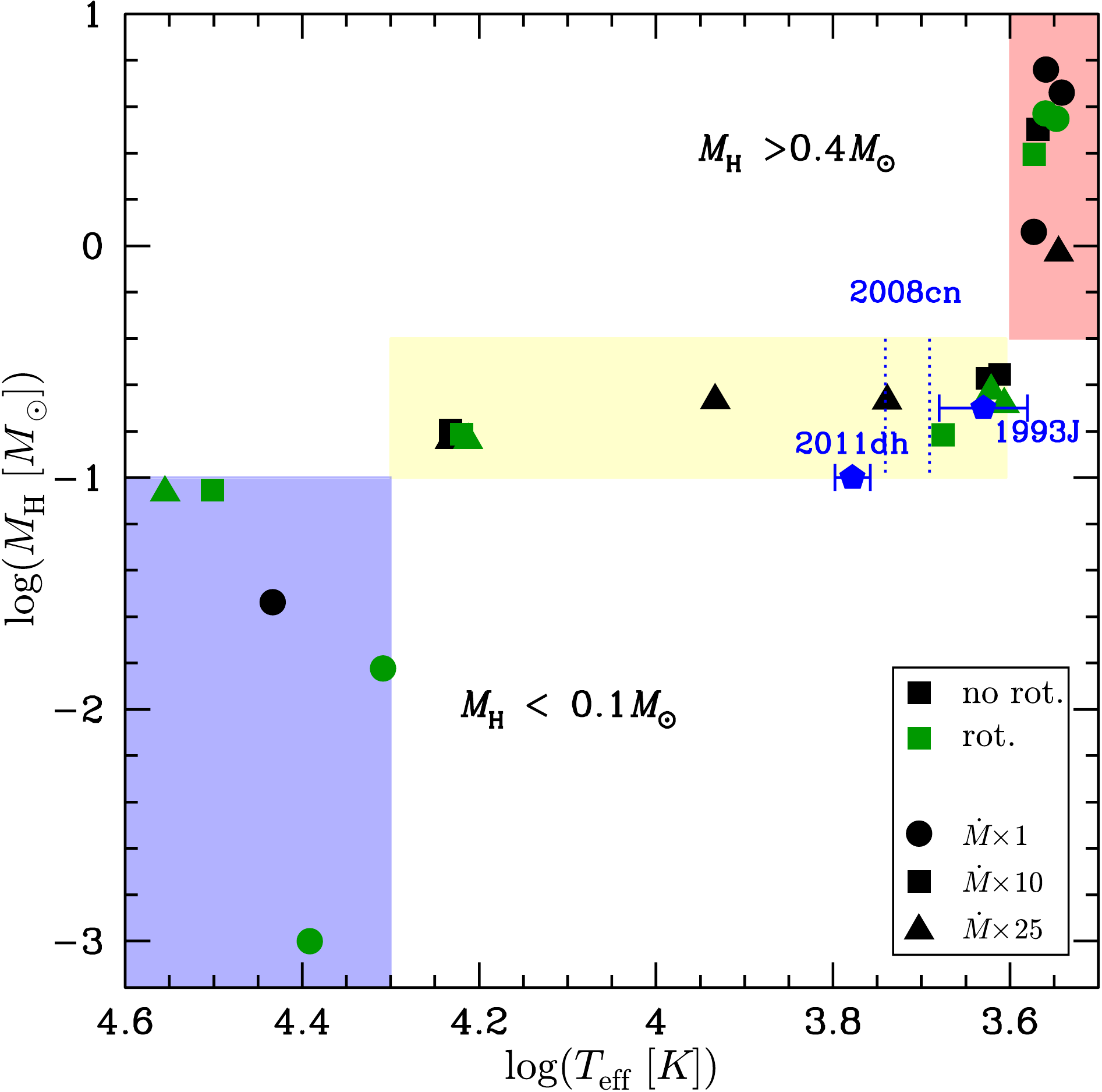}
   \caption{Mass of hydrogen in solar masses at the pre-supernova stage for the various models of Table 1. Positions in this diagram of some supernovae are indicated by pentagons with error bars. In the case of the SN2008cn, the mass of hydrogen is not known, the range of effective temperatures is framed by the two dotted vertical segments. The three shaded regions from left to right correspond to stars ending their lifetime as blue, yellow and red supergiants.}
      \label{henv}
\end{figure}

Figure~\ref{COHEf} shows the structure of the pre-supernovae models for different initial masses, rotation and prescriptions for the mass-loss rate during the RSG phase.
In Table~\ref{Tab2}, the final radii, masses, masses of the hydrogen-rich envelope\footnote{The mass of the hydrogen-rich envelope, $\Delta M_{\rm H}$, is simply the difference
between the final mass and $M_\alpha$. Note that hydrogen can also be present outside $\Delta M_{\rm H}$, in the outer layers of $M_\alpha$.  This is the reason why, in some models, the integrated mass of hydrogen, $m_{\rm H}$, can be larger than $\Delta M_{\rm H}$. Here, $M_\alpha$ is defined here as the
lagrangian mass coordinate where the mass fraction of helium becomes superior to 75\% going from the surface to the center.}, the masses of the helium, carbon-oxygen cores, of the remnants are indicated together with the integrated mass in the envelope of the quantities of hydrogen, helium, carbon and oxygen.
For the rotating models, the angular momentum in the remnant, the angular velocity of the neutron star and its period at birth are also provided. For computing these quantities we followed the same method as in \citet{Georgy2012WR}.
We can note the following effects of enhancing the mass-loss rate during the RSG phase:
\begin{itemize}
\item From table~\ref{Tab2}, we see that the models with enhanced mass-loss rates during the RSG phase have in general smaller He and CO cores at the end of their evolution, which is of course expected. We note however that due to the interplay between mass-loss and lifetime during the RSG stage, some models may present slightly larger He or even CO cores for higher RSG mass-loss rates (see for instance the case of the rotating 20 M$_\odot$ with 1 and 10 $\times$ the standard mass-loss rate). On the whole, however, the effects on the cores  remain modest. 
\item We see also that in the most extreme case of mass-loss during the RSG phase considered here, the final mass of the star
contains 30-40\% of the initial mass, whatever the initial mass between 9 and 25 M$_\odot$ (see the bottom panels of Fig.~\ref{COHEf} and, in these panels, the bottom line
framing the ``WIND'' region).
\item One can wonder whether the change of the mass-loss rate during the RSG phase can impact the angular momentum of the core? 
We see that the angular momentum in the remnant decreases when the RSG mass-loss rate increases. However, the changes are very modest. Even considering the models with a mass-loss
increased by a factor 25 would produce extremely rapidly rotating neutron stars. For instance the largest period obtained here for a neutron star would be of 3.6 ms. 
For comparisons, the observed shortest periods for young pulsars are around 20 ms, thus five time larger than the periods obtained here. 
So some angular momentum should still be lost, either during the previous phases \citep{Heger2005} or at the time of the SN explosion \citep{Blondin2007} or during the early phases of the evolution of the new born neutron star. 
\item The RSG mass-loss increase has the strongest impact on the structure of the envelope in the range of initial masses between 9 and 15 M$_\odot$.
We see indeed that very little changes occur for the 25 M$_\odot$, while important changes
occur for the 9 or 15 M$_\odot$. This is quite natural since, the 25 M$_\odot$ models 
spend a very short time into the red supergiant phase anyway, so that changing the mass-loss rates during the short RSG phase will have only a marginal impact. 
This also justifies the reason why we stopped our
investigation to this upper mass limit. 
\item We see that the H-rich envelope is significantly reduced by the high RSG mass-loss rates. On the other hand, as already noted above,
the masses of the He-rich layer and of the CO core are generally only slightly changed.
\item Looking at the structure of stars finishing their life as red supergiants (see red stars in the upper part of each panel of Fig.~\ref{COHEf}),
we  note that red supergiants can exist for very different masses of the H-rich envelope \citep[see also][]{Groh2013}. Actually the range for the masses of H-rich envelopes in RSG pre-supernova models can range from
more than 80\% the initial mass down to only a few percents of the total initial mass!  For the models with $\log T_{\rm eff}> 4.3$, they are likely WR stars \citep{Groh2013} and the mass of the H-rich envelope covers a much more restricted range from 0.004\% to a maximum of 1\%. The pre-supernovae models with intermediate colors between the red and the blue (these are LBVs or YHGs at the pre-explosion; \citealt{Groh2013}) present H-rich envelope covering the range between about 1 and 5\%.
So we see that whenever the H-rich envelope contains more than about 5\% of the initial mass, the star will end as a red supergiant, and whenever the whole H-rich envelope is less than 1\% of the total mass
the star appears as a WR star before the SN. For intermediate situations, intermediate colors/effective temperatures are obtained and the star appears as an LBV or YHG at the pre-explosion stage.
\item Instead of looking at the mass  of the H-rich envelope, we can look at the integrated mass of hydrogen in the pre-supernova models. 
Fig.~\ref{henv} shows the mass of hydrogen in the star at the pre-SN stage as a function of the effective temperature. All stars with a mass of hydrogen above 0.4 M$_\odot$ are 
red supergiants, all stars with a mass of hydrogen inferior to 0.1 M$_\odot$ appear as WR stars. Stars with intermediate values  appear as LBVs/YHGs with effective temperatures between those corresponding to red supergiants and late-type WR stars.
%If we use instead of the mass of hydrogen, the ratio of the mass of hydrogen to that of helium, we see that  the progenitors of core collapse supernovae having
%colors intermediate between the red and the blue should present ratios of the mass of hydrogen to helium between 0.05 and 0.15. 
\item For those stars that have intermediate $T_{\rm eff}$ (LBVs and YHGs), we can wonder what is the parameter which governs the effective temperature or the total radius of the model.
In Table \ref{sec} below, we have indicated
for the models belonging to the yellow region of Fig.~\ref{henv}  various properties. Models are ordered from top to bottom by increasing effective temperatures.
We see that, in general, the effective temperature increases when the actual mass, the mass of the CO, or of the He core increases.
Also, generally, the effective temperature increases when the mass of hydrogen in the envelope decreases, or when the ratio of the mass of hydrogen to that of helium decreases.
\end{itemize}

\begin{table}
\caption{Properties of the last computed models with effective temperatures between 3.6 and 4.3 ordered
from top to bottom by increasing effective temperatures (in logarithm). The luminosity is in logarithm, the masses in solar masses. The model nn/mm/pp corresponds
to the models with an initial mass nn, a RSG mass-loss rate increased by a factor mm, and an initial rotation on the ZAMS equal to pp times the critical velocity $\upsilon_{\rm crit}$.}
\label{sec}
\scriptsize{
\centering
\begin{tabular}{ccccccccc}
\hline\hline
  T$_{\rm eff}$    &  L/L$_\odot$  & Model      &  M$_{\rm fin}$ & M$_\alpha$ & M$_{\rm CO}$ &  ${L_{\rm H} \over L_{\rm core}}$ & m$_{\rm H}$ & ${m_{\rm H}\over m_{\rm He}}$ \\
\hline
3.606                        &  4.057                    &  9/25/0.4   &  2.60                & 2.27            &  0.94             &     1.44                                   & 0.21            & 0.16                                                 \\ 
3.612                        &  4.856                    &  15/10/0    &  4.88                & 4.24            &  2.24             &     1.58                                   & 0.28            & 0.14                                                \\ 
3.621                        &  4.926                    &  15/25/0.4 &  5.32                & 4.70            &  2.71             &     1.70                                   & 0.24            & 0.12                                                \\ 
3.626                        &  5.039                    &  20/10/0    &  6.68                & 6.10            &  3.81             &     1.27                                   & 0.27            & 0.11                                                \\ 
3.655                        &  4.944                    &  15/10/0.4 &  5.31                & 4.92            &  2.79             &     1.60                                   & 0.15            & 0.08                                                \\ 
3.739                        &  4.787                    &  15/25/0    &  4.69                & 4.70            &  2.21             &     1.55                                   & 0.21            & 0.11                                                \\ 
3.933                        &  5.027                    &  20/25/0    &  6.50                & 6.04            &  3.75             &     1.23                                   & 0.22            & 0.09                                                \\ 
4.212                        &  5.178                    &  20/25/0.4 &  7.26                & 6.89            &  4.52             &     1.26                                   & 0.14            & 0.07                                                \\ 
4.219                        &  5.181                    &  20/10/0.4 &  7.36                & 6.96            &  4.86             &     1.27                                   & 0.15            & 0.07                                                \\ 
4.232                        &  5.221                    &  25/10/0    &  8.26                & 7.90            &  5.54             &     1.11                                   & 0.16            & 0.07                                                \\ 
4.233                        &  5.262                    &  25/25/0    &  8.35                & 8.02            &  5.66             &     1.11                                   & 0.14            & 0.06                                                \\ 
 \hline
\end{tabular}
}
\end{table}

Let us end this section by saying a few words about the surface rotational velocities of post RSG stars. There is a big difference between the case of the 9 M$_\odot$ model and the more massive models considered here.
The rotation velocity of the 9 M$_\odot$ is higher along the blue loop than during the first crossing of the HR gap (standard mass loss rate model) . Typically at an effective temperature of Log $(T_{\rm eff}/K)=3.8$, the 9 M$_\odot$ model has a surface velocity  of about 18 km s$^{-1}$ during its first crossing. This velocity becomes about 106 km s$^{-1}$ during the second crossing and nearly 43 km s$^{-1}$ during the third crossing ! In that case, we see that from the RSG stage where the surface velocity is very small, we have a rapid increase of the surface velocity when the star contracts to the blue. Now if we consider the case of the 15 M$_\odot$ model (model with 10 times the standard mass-loss rate), we have a very different situation. At Log $(T_{\rm eff}/K)=3.8$, the surface rotational velocity is between 9 and 10 km s$^{-1}$ during the first crossing, and only 0.3  and 0. 1km s$^{-1}$ during the second and third crossing.
This very different behavior results from the fact that in the 9 M$_\odot$ model, the blue ward evolution results from the mirror effect (see section 2), while for the 15 M$_\odot$, as well as for more massive stars, the blue ward evolution
results from strong mass-loss. In the case of the 9 M$_\odot$ model, one has that angular momentum is dredged up to the surface by the deep convective zone during the RSG phase. This model does not lose  a lot of mass (see Table 1), therefore this angular momentum remains in the star and produces an acceleration of the envelope when the star contracts to the blue \citep{HL1998}. In the case of the 15 M$_\odot$, the evolution to the blue part of the HR diagram is due to strong mass-losses which remove also a lot of angular moment making the surface velocities very low during the post RSG stages. So interestingly, the physical mechanism responsible for the blue ward evolution has an important impact on the surface velocities in the yellow and blue supergiant domain.

%voir HRD avec positions WR

%In contrast, the rotating models, all end their lifetime as a WR stars. GIVE HERE LIFETIME DURING WHICH STAR IS A WR STAR. Higher the has loss rate during the RSG phase, longer will the WR phase. EXPLAIN HYD IN 1X MODEL. DISCUSS THE NON OBSERVATION OF SIMULTANEOUSLY THE WR AND RSG IN SINGLE AGE POPULATIONS. DOES THIS DISCARD
%THE STRONG mass-loss MODELS? NO ONE OF THESE STARS DO END AS A WC STAR! THIS KIND OF SCENARIO CANNOT BE INVOKED TO
%EXPLAIN THE LOW LUMINOUS WC STARS (EVEN WHEN HIGH MDOT IN RSG COMES FROM MASS TRANSFER). COULD THESE STARS BE DUE TO
%MASS TRANSFER IN CLOSE BINARY IN OTHER PHASE? (I DO NOT THINK SO SEE RESULTS BY AUDREY, LIKELY ANOTHER PAPER WITH NEW BINARY MODELS, SONG MODELS?).

%provide orders of magnitudes.

\subsection{Core collapse supernovae with a post-red supergiant progenitor}

Some supernovae are observed to have a yellow or a blue progenitor while being in the mass range of stars evolving into a red supergiant phase during their lifetime. 
Most likely these stars reach that point because they lost significant amounts of mass during the red supergiant phase. 
We present in Table~\ref{TableIIYSG}, the list of known progenitors of type II supernovae for which the progenitors was not a red supergiant.
Their positions in the HRD diagram are shown in Figs.~\ref{snrsg}. 

Obviously, the non-rotating models with normal RSG mass-loss rate cannot account for the existence of most if not all those stars. 
All the observed yellow progenitors can be explained in the frame of the non-rotating models computed with some enhancement of the RSG mass-loss rates. 
However rotating models, with standard mass-loss rate,
present blue ward evolution above a luminosity around 5.0, thus these models may be also invoked to explain some post red supergiants progenitors with luminosities above 
this limit\footnote{Note that we have computed only two sets of models with different rotational velocities (0 and 0.4$\upsilon_{\rm crit}$). In reality, we have a distribution of initial rotations and thus the domain of SN progenitors covered by the rotating models is larger than the one shown in the right panel of  Fig.~\ref{snrsg}.}. Thus there is some degeneracy between the effects of rotation and those induced by different RSG mass-loss rates. Is there any possibility to discriminate between these two possibilities by observing yellow supergiants?
The surface velocities in these stages are very low, so difficult to measure and moreover, their values tell very little about the initial rotation. The surface abundances also do not appear very constraining.
May be more hope can come from asteroseismology. The possibility to probe for instance the internal rotation could may be a way to differentiate between rotation or mass loss as the main cause for the existence of these yellow progenitors.
%When mass-loss rates are enhanced during the RSG phase, it is possible to reproduce the positions of these supernovae, even those with luminosities below 5.0.
 
Let us now discuss each of the 6 supernovae identified as having a yellow supergiant as progenitor.
%provide estimate of the range of initial masses for the various progenitors.
%The luminosities of SNe IIP are distributed in a broad range (luminous, normal, subluminous and faint events). Not clear what determines the difference. Progenitor mass (Utrobin \& Chugai 2011), higher luminosities from higher masses.

\begin{table}
\caption{Properties of post-RSG identified as SN progenitor. Column 1 gives the number used to label the points in Figs.~\ref{snrsg}.}
\label{TabWind}
\centering
\scriptsize{
\begin{tabular}{ccccccc}
\hline\hline
 &    SN   & type & [O/H] & Log L/L$_\odot$  & Log T$_{\rm eff}$ & Reference \\
\hline
1& 1993J  & II-Ib        &  -                          &  5.1$\pm$0.3         & 3.63$\pm$0.05     & [1] \\   
2& 2008cn & II-P        & 8.76$\pm$0.24   & 4.93$\pm$0.1        & 3.716$\pm$0.025 & [2] \\
3& 2009hd & II-L       &  8.43$\pm$0.05   & $\le$5.04               & $\le$ 3.725            & [3] \\
4& 2009kr & II-L        &  8.67                     & 5.12$\pm$0.15      &  3.724$\pm$0.045 & [4] \\
 &            & IIn-II-P  & 8.06$\pm$0.24    & 5.1$\pm$0.24        &  3.685                    & [5]  \\  
5& 2011dh &   IIb       &  $\sim$8.7            &  4.92$\pm$0.20     & 3.778$\pm$0.02 &  [6] \\   
 &             &   IIb       &  $\sim$8.7            &  4.99                      & 3.778$\pm$0.01 &  [7] \\     
6& 2013df  &   IIb       &   $\sim$8.7            &  4.94$\pm$0.06    & 3.628$\pm$0.01 & [8]  \\                     
\hline
\end{tabular}
\label{TableIIYSG}
\tablefoot{[1]=\citet{SN93J2004}; [2]=\citet{SN08cn2009}; [3]=\citet{SN09hd2011}; [4]=\citet{SN09kr2010}; [5]=\citet{SN09krF2010}; [6]=\citet{SN11dh2011}; [7]=\citet{SN11dhV2011}; [8]=\citet{SN13df2014}\\}
}
\end{table}

{\it SN 1993J}: this supernova originated very likely from a close binary system, consisting, a few thousand years before explosion, of a red (the primary) and a blue supergiant. Roche Lobe Overflow from the
red supergiant caused its blueward evolution \citep{Pod1993,SN93J2004}. This scenario is strongly supported by the detection of the hot component of this close binary system \citep{SN93J2004}.
Although the present models are for single stars, the enhanced mass-loss rate models can somewhat mimic the effect of a Roche Lobe Overflow. We see that the rotating 15 M$_\odot$ model with a mass-loss increased by a factor 25 during the red supergiant stage would provide
a reasonable fit to the progenitor of 1993J. The luminosity is actually 0.2 dex below the  attributed luminosity but still in the error bar. Interestingly the lightcurve is well matched with models of an explosion of a He-core of mass
4-5 M$_\odot$ which had a low mass H-envelope of around 0.2 M$_\odot$ \citep{Nomoto1993, Pod1993, Woosley1994}. 
Using the values of the effective temperature determined by \citet{SN93J2004} as well
as the mass of hydrogen in the envelope we can place the position of this SN progenitor in Fig.~\ref{henv}. 
Our rotating 15 M$_\odot$ with 25 times the standard
mass-loss ends its lifetime with 5.3 M$_\odot$ and a low mass H-envelope of 0.2 M$_\odot$, the He-mass in the pre-supernova model is 1.9 M$_\odot$, its CO core mass is 2.2 M$_\odot$. 
So this model would provide a good fit not only to the observed position
in the HR diagram but likely also for the evolution of the supernova lightcurve.
%Moreover it provides a prediction for the mass of helium in the ejecta (the mass of heavier species are affected by the mass cut while that of helium is not).

{\it SN 2008cn}: A progenitor candidate has been proposed by \citet{SN08cn2009}. Its yellow color (see the right panel of Fig.~\ref{snrsg}) would place it among the yellow supergiants. According to \citet{SN08cn2009}, it might be that the yellow progenitor could arise from the blend of two or more stars, such as a red supergiant and a brighter, blue supergiant. 
Actually, the fact that the light curve did appear as a type II-P plateau would favor the
explosion of a red supergiant instead of a yellow one.  
Actually, if we take for granted that the progenitor was the yellow supergiant, then from comparisons with the present models we can deduce the
following properties: a non-rotating 18 M$_\odot$ model with mass-loss increased by more than 10 times the standard mass-loss during the RSG phase could likely provide a good fit to the observed position 
of the progenitor. A rotating progenitor with masses between 15-17 M$_\odot$ with an increased mass-loss during the RSG phase (between 10 and 25 times the standard one) would likely provide a reasonable
solution too. This would mean that the actual mass of the progenitor would be between 5-8 M$_\odot$, the mass of ejected hydrogen around 0.15-0.20 M$_\odot$ and that of helium around 1.9 M$_\odot$. 
%Actually such low mass of hydrogen is likely difficult to reconcile with a type IIP supernova.

{\it SN 2009hd}: This object is heavily obscured by dust. Via insertion of artificial stars into the pre-SN HST images, \citet{SN09hd2011} could constrain the progenitor's properties. The magnitude
and color limits are compatible with a luminous red supergiant, they also allow for the possibility that the star could have been more yellow than red. Actually the point put in Figs.~\ref{snrsg} represent
the upper values for the luminosity and effective temperature. These limits are very similar to that attributed to SN 2008cn and therefore the same estimates concerning the actual mass, the masses of
H and He ejected can be made (see Sect.~5.1). 
%Let us note that the fact that here we have a type II-L supernova event makes more reasonable the fact that this SN may have had a yellow supergiant progenitor.

{\it SN 2009kr}: Properties of the progenitors have been obtained by two teams, \citet{SN09kr2010} and \citet{SN09krF2010}. The properties obtained by the two teams are in relatively good agreement. The only point
on which a large difference exists is on the metallicity inferred for the region where the supernova occurred (see Table \ref{TableIIYSG}). This illustrates the fact that indeed metallicity estimates are at the moment difficult
and not very reliable. Another point where the discussion between the two teams differs is on the SN type. \citet{SN09kr2010} from their own analysis conclude that SN 2009kr is a type II-L supernova, while \citet{SN09krF2010}
reports that \citet{Tendulkar2009} indicated that SN 2009kr showed the features of a II-n SN. They also report that \citet{Steele2009} claimed SN 2009kr be a type II-P. 
In view of the inferred position in the HR diagram and
the low mass of H that it implies, we tend to support the conclusion by  \citet{SN09kr2010} that we have here a type II-L SN event. As for 2009kr, positions of the progenitor in the HR diagram support the view
that the star is a post red supergiant star having lost a great part of its envelope as proposed by \citet{Georgy2012}.

{\it SN 2011dh}: this supernova attracted much attention, and discussion about the nature of its progenitor has been recently resolved. Actually until recently it was uncertain whether the progenitor of this supernova was
a compact or an extended star. Some authors \citep{Arcavi2011}, based on the properties of the early light curve and spectroscopy, suggested that the progenitor was a member of the compact IIb family
\citep{Che2010} and that the progenitor identified by \citet{SN11dh2011} was actually not the progenitor but possibly a companion to the progenitor or a blended source, as its radius ($\sim$ 10$^{13}$ cm) would be highly inconsistent with constraints from their post-explosion photometric and spectroscopic data.  On the other hand, \citet{Bersten2012} used a set of hydrodynamical models to study the nature of the progenitor of SN 2011dh. Their modeling suggests that a large progenitor star, with a radius about 200 R$_\odot$ is needed to reproduce the early light curve. Their model would thus support the identification of the progenitor with the yellow supergiant
detected at the location of the SN event in pre-explosion images. Their model gives a mass of the ejecta to be around 2 M$_\odot$, the progenitor was composed of a helium core of 3-4 M$_\odot$ and a thin
hydrogen-rich envelope of about 0.1 M$_\odot$. Actually, \citet{VAN11dh2013} have shown using HST observations of the regions of SN 2011dh about 641 days after the explosion, that the yellow supergiant has disappeared
implying that this star was the progenitor of the SN. Recently  a candidate for the companion of the progenitor of this supernova has been detected by \citet{Folatelli014}.
From the models presented here we see that whatever cause the mass-loss, it should have been quite important to make the star to evolve at that position at the time of the supernova explosion.
Actually the progenitor position would be compatible with an initial mass around 15-18 M$_\odot$ having undergone more than 10 times the standard mass-loss during the red supergiant stage. Models
with rotation would favor an initial mass around 15 M$_\odot$, while non-rotating ones would favor a higher initial mass around 18 M$_\odot$. Interestingly again, non rotating as well as rotating models would predict an actual mass at that position
between 4 and 6 M$_\odot$, with an hydrogen mass in the envelope around 0.2 M$_\odot$, so not so far from the estimates made by \citet{Bersten2012} on the basis of the early light curve and spectra of the SN.

{\it SN 2013df}: \citet{SN13df2014}  analyzing archived observations of the HST obtained 14 years prior to explosion, have identified the progenitor to be a yellow supergiant. This supernova shows some similarities with SN 1993J, although with less $^{56}$Ni ejected and with a progenitor more extended in radius. It is clearly of type IIb. 

For some of these SNe, the enhanced mass-loss models (both rotating and non-rotating) can provide a reasonable fit not only to the observed position in the HRD but also to the masses of the progenitors and of the hydrogen and or helium as they can be deduced from the observed properties of the SN light curve. 
So these post RSG progenitors  provide some support to the enhanced mass-loss rate models, while, as we saw in the previous section, the RSG progenitors favors more the standard mass-loss rate models. 

This implies that a star with a given mass and metallicity may go through different mass-loss regimes during the RSG phase. What triggers the different mass-loss regimes may be the presence of a close companion, rotation, the presence of a magnetic field... Actually the origin of the differences are not lacking, the question which remains open is to identify them and to estimate their frequency.
%but to proceed
%we need as a first step to have a better idea of the frequency of these post RSG objects. We further discuss this point below in the perspectives.

\section{Mass-loss rates and populations of evolved stars }

%WR
%discuter table
% discuter rapport B/R, Y/R, WR/R
% Comparer RSG/O avec amas.

\begin{figure}
%: fig TcrcZ014_Mcc_rot.eps
\centering
\includegraphics[width=.49\textwidth]{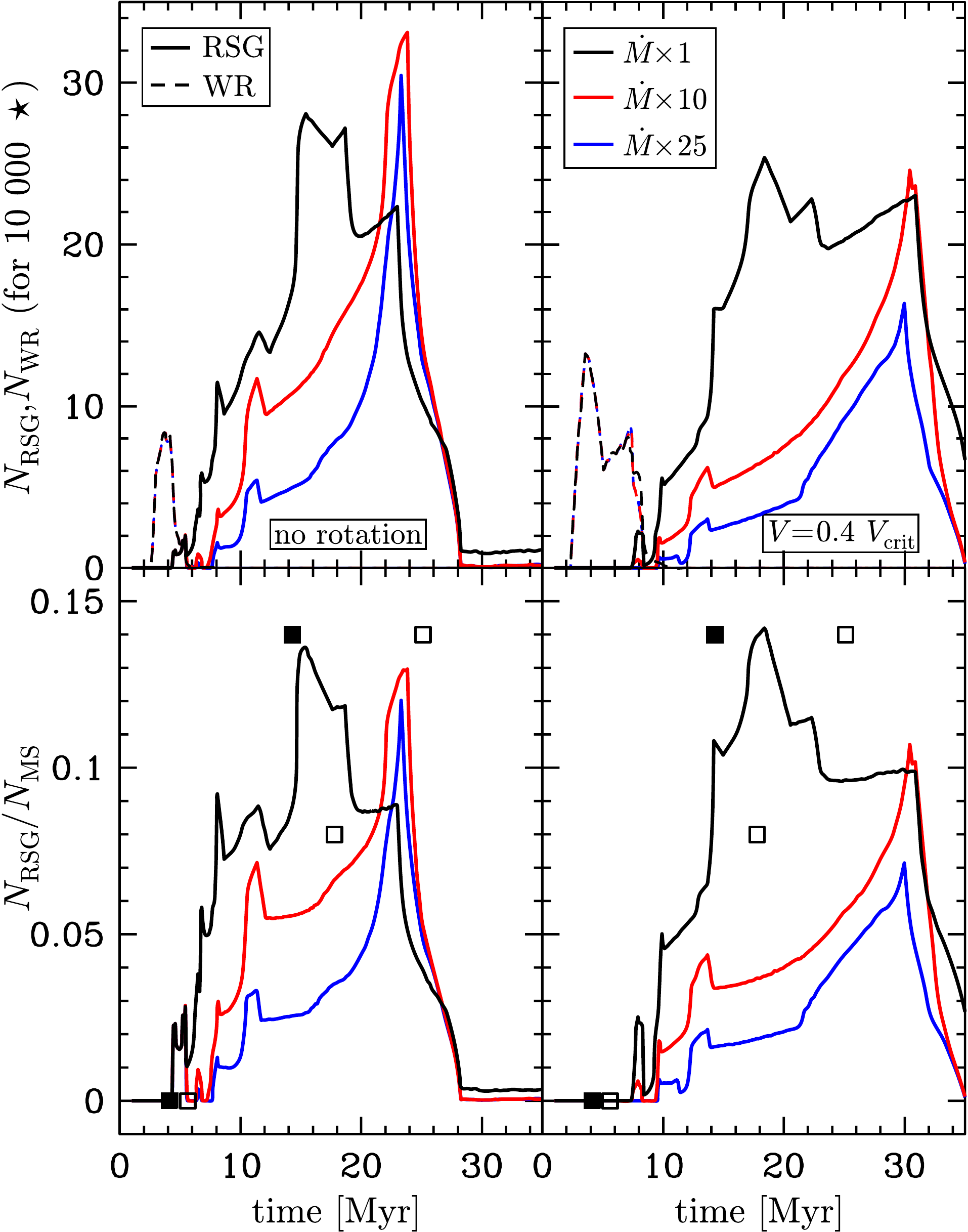}
   \caption{{\it Upper panels:} Expected number of Wolf-Rayet stars (dashed lines) and of Red Supergiants in a coeval population (continuous lines) of 10000 stars. Models with initial masses between 9 and 25 M$_\odot$ computed with different mass-loss rates during the Red Supergiant phases  give the different continuous lines labeled by the enhancement factor considered for the mass-loss rate during the RSG phase. 
   The
   populations of WR stars are very little affected by these changes of mass-loss recipes (see text). 
   {\it Lower panels:} Fraction between the number of RSG and the number of Main-Sequence stars two magnitudes below the turn off. The black and empty squares correspond to observed ratios in stellar clusters with respectively between 20 and 50 stars and with less than 20 stars. From left to right, the points correspond to the following clusters: NGC 1976, NGC 6231, NGC 884+4755, NGC 457, NGC 581+2439 \citep[see][]{Meynet1993}.}
      \label{POPU}
\end{figure}

%RSG
When the mass-loss rate is increased during the RSG phase, we expect to observe in general a smaller number of RSG stars since the RSG lifetimes are decreased. Moreover the upper mass limit of stars spending
some time in the RSG stage is decreased. This can be seen in the upper panels of Fig.~\ref{POPU}. 
We see that for the standard mass-loss rates and no rotation (upper left panel), the number of red supergiants varies between 1 and 28 in a cluster of 10 000 stars with
ages between about 5 and
28 Myr. When a mass-loss rate one order of magnitude greater is used during the RSG phase, we see first that
red supergiants appear later than in the case of standard mass-loss, the first red supergiants appear around 8 Myr instead of 4 Myr\footnote{One can wonder why changing the mass-loss rate only during the RSG phase can modify the time of appearance of the first supernovae after an instantaneous burst of star formation. The code is computing the upper mass limit of stars becoming RSGs by computing through extrapolation of the durations of the RSG phases, the lower initial mass model having a RSG lifetime equal to zero.  This procedure implies that when stronger RSG mass-loss rates are used, since the RSG lifetimes are reduced, this limit is lowered and thus
correspond to larger ages.}. The number expected are smaller than for the models computed with a standard mass-loss rate during the whole period between 8 and about 20 Myr. The case with an increased factor of 25 shows a qualitatively similar behavior but with a stronger decrease during the period between 8 and 20 Myr.

%RSG-MS
The two lower panels compare the predicted number ratios of RSG with MS stars (two magnitudes below the turn off) with the observed numbers in a few  open clusters having ages between 4  and 25 Myr. 
%Note that the ages considered here are those obtained by isochrone fitting of the old Geneva tracks (REF). Rigourously one should have used here the ages obtained by isochrone fitting using the present models.
%However this will produce only slight differences by at most 
%10-20\% in ages and thus will not have a big impact on the conclusions that can be drawn from this comparison. 
A point to keep in mind is the fact that due to the small number of stars, stochastic effects can largely blur the picture. 
We see that the two youngest clusters do not show any RSG populations. This may be compatible with the rotating models. 
The other clusters present RSG/MS ratios between 8 and 14\%. These values do appear more compatible with the standard mass-loss rate models than with the increased mass-loss rates ones.
Although this should be seen at the moment as a weak constraint, we can just retain at the moment that
the present situation would favor, for the bulk of the RSG populations, a time-averaged mass-loss rate compatible with the standard mass-losses used by the \citet{Ekstrom2012} models. 
%This of course cannot discard the possibility that some RSG will undergo stronger mass-loss rates but at least this first
%comparison would eliminate the possibility that the bulk of the RSG would undergo increased RSG mass-loss rates.

%RSG/BSG/YSG
Enhancing the RSG mass-loss rate increases the number ratio of blue to red supergiants.
Typically, for the non-rotating (rotating) 15 M$_\odot$ model, the ratio B/R (=$(t_{B1}+ t_{B2})/t_{\rm RSG}$) passes from a value equal to 0.20 (0.05) for the standard model to 0.3 (6.8) and 1.7 (6.9) for 
respectively the 10 and 25 $\times$ the RSG mass-loss rate models. 
The number of blue supergiants observed in solar metallicity clusters with mass at the turn off around 9-15 M$_\odot$ \citep{Meylan1983, Eggenberger2002} is around 2. We see therefore
that, if a fraction of the 15 M$_\odot$ stars would undergo stronger mass-losses during the RSG phase, then it would help in making the theoretical ratio compatible with the observed ones. However this question deserves more studies for the following reasons: 1) in the case of the 9 M$_\odot$, enhancing the mass-loss rate will actually decrease the  $(t_{B1}+ t_{B2})/t_{\rm RSG}$ ratio through the suppression of the blue loop; 2) we know that
the enhanced mass-loss rates will likely occur only for a fraction of stars, since some core collapse progenitors are red supergiants which cannot be reproduced by RSG enhanced mass loss rate models and, as seen above, the number ratio of RSG to MS stars do appear to be better fitted by the standard mass-loss rate models; 3) finally, studies at other metallicities should be undertaken since the real challenge is not to reproduce
the B/R at a given metallicity but to reproduce the general trend which is an increase of the ratio B/R with metallicity.

%RSG-WR
As we have shown, models with enhanced RSG mass-loss rate (and no other further changes as enhanced post RSG mass-loss rates) do not produce WC stars. 
Our models indicate that simultaneous presence of RSG and WR stars of the WN type would only occur for a very limited age range, indicating that
single-aged populations showing these two population should be relatively seldom. 
%This is for the moment in good agreement with the observations ({\bf WE NEED REFERENCES HERE}).
There are indeed not many cases of single aged populations showing both red supergiants and Wolf-Rayet stars. We can cite Westerlund 1 \citep[][although it is so massive that not all the stars may be coeval]{Clark2005}
and clusters in the centre of the Galaxy  \citep{Figer2004, Figer2009}. 
%{\bf In addition, a fraction of stars evolve through the binary channel, which modifies the ratio WR/RSG (WE NEED A CITATION HERE).}
Interestingly \citet{Shara2013}, studying 
the massive star population in the ScI spiral galaxy M101 found that the spatial distributions of the WR and RSG stars near a giant
star-forming complex are strikingly different. WR stars dominate the complex
core, while RSG dominate the complex halo. In case this difference is linked to an age difference, it could be explained by the fact that these two populations originate from
different initial mass ranges.

We mentioned above that actually models with enhanced mass-loss rates decrease the number of WR star formed (see Table~\ref{TabBR}). However the mass range considered here between 9 and 25 M$_\odot$ contributes very little to the WR population that, in the present models, come mainly from  more massive stars. This is why in Fig.~\ref{POPU}, we see no difference for what concerns the WR populations between the different models.

% PLUS LOIN

%The question of how to explain  the observed variations with the metallicity of the number of blue to red supergiants  is still an open question (REF). Clearly, mass-loss and mixing play a role, may be binary also. 

\section{Conclusions and perspectives}

Red supergiants can be the end point of the evolution of massive stars or a transition phase before the star evolves into bluer parts of the HR diagram exploding when it is a yellow, a blue or even a  WN-type Wolf-Rayet star. 
In this work, we explored how a change of the mass-loss rates during the RSG phase influences on one side the properties of the RSG and on the other side the evolution during  the post RSG phases and the nature of the
core collapse progenitors. The present work has reached the following conclusions:
\begin{enumerate}
\item For the RSG, the positions in the HRD, radii, surface abundances and rotation velocities are mostly insensitive to the RSG mass-loss rate used.
\item  Adopting different mass loss rates during the RSG phase changes the initial mass attributed at a given red supergiant by evolutionary tracks. When enhanced mass loss rates are used, in general, higher initial masses are
associated to a given RSG star.
\item Since the mass to be lost for a given initial mass star to leave the RSG phase is more or less fixed, enhancing the RSG mass-loss rates beyond the point making a blue ward evolution to appear, does not change
the total mass lost during the RSG phase, but the duration of the RSG phase.
For a star with an initial mass between 15 and 25 M$_\odot$, the maximum mass that
can be lost during the RSG phase is between 40-60\% of the initial mass.
\item A consequence of the link between RSG lifetime and mass-loss rate is that a change of the RSG mass-loss rate has a strong impact on the luminosity function of RSGs.  
\item At solar metallicity, the enhanced mass-loss rate models do produce significant changes on the populations of blue, yellow and red supergiants. When extended blue loops are produced by enhanced mass-loss, the models predict that a majority of blue (yellow) supergiants are post RSG objects\footnote{One exception however is for the rotating 20 M$_\odot$ for which the duration of the yellow supergiant phase
before the RSG phase is longer than the duration of the same phase after the RSG phase.}
\item Enhanced mass-loss rates during the red supergiant phase has little impact of the WR populations. We can safely say that even the enhanced mass-loss rate models cannot reproduce the low luminosity WC stars. As indicated above post-RSG mass-loss rates during LBV's pause could help.
In a work in preparation, we investigate whether these stars can be produced in close binary systems, with mass transfer occurring before the RSG phase, but this does not appear as a promiseful channel either for explaining those stars (Barblan et al. in preparation). Another possible solution would be that those stars come from massive stars with higher mass-loss rates \citep[like in the models discussed by][]{MM1994}.
\item We show that the position in the HRD of  the end point of the evolution depends on the mass of the hydrogen envelope, a point already emphasized in \citet{Groh2013}. More precisely, 
whenever the H-rich envelope contains more than 5\% of the initial mass (actually the limit value may be between 5 and 10\%) , the star will end as a red supergiant, and whenever the whole H-rich envelope is less than 1\% of the total mass, the star is a blue supergiant. For intermediate situations, intermediate colors/effective temperatures are obtained.
\item An enhanced mass-loss rate during the RSG has some impact on the angular momentum of the core, but at a level which is much too low to allow this effect to be invoked for 
explaining the observed rotation  rate of pulsars. This conclusion holds in case no other coupling than those induced by shear and meridional currents exists. In case a strong coupling would be active
(due for instance to a strong magnetic coupling), then things can be very different.
%However, according to our models, see point above, stars evolving back to the blue  likely does not constitute the bulk of stars evolving into a RSG phase.
\item Concerning the question, what are the RSG mass-loss rates which are favored by observations, we can bring the following elements of response: first, RSG mass-loss rates deduced from spectroscopy show a very large scatter
and an outburst behavior for the mass-loss rates of RSGs.  Likely, the mass lost by a star with a given initial mass and metallicity may be different depending on some additional characteristics of the
star (star may have a close companion, rotation\footnote{In the present models we have explored some effects of rotation but, for instance, the impact of rotation on the pulsation properties of red supergiants
has not been studied yet. Also a strong coupling due to an internal magnetic field may bring some diversity in the evolutionary scenarios}, ...). 
Arguments based on the positions of the red supergiant progenitors of type II-P supernovae 
%The transition between red supergiants being the end point of the evolution and red supergiants being in a transitory phase could be around 17 M$_\odot$ at solar metallicity. This would explain why Smartt (2009). 
and on the RSG populations in clusters indicate that
the standard mass-loss rates are likely appropriate for describing the bulk of the RSG populations. On the other hand, the existence of yellow or blue progenitors with initial masses between 15 and 25 M$_\odot$ favors RSG enhanced mass-loss rates. Interestingly, \citet{Humphreys2013} identified stars in the galaxies M31 and M33 that they call warm hyper giants which present properties favoring a post RSG phase.  These stars thus likely formed through enhanced RSG (or may be post RSG) mass-loss rates. The properties of warm hyper giants are A to F-type absorption spectra, winds with relatively slow outflows, an extensive and dusty circumstellar ejecta, and relatively high mass-loss. 
The warm hypergiants show the small oscillations often referred to as $\alpha$ Cygni variability \citep{vanGenderen2002}. This would be exactly in line with the conclusion by \citet{Saio2013}, who explain the 
pulsational properties of the alpha Cygni variables, which are blue supergiants with masses around 20-25 M$_\odot$, by the fact that they are post RSG stars \citep[see also the discussion in ][]{Georgy2013}. 
More work is needed to know the main physical cause(s) for enhanced RSG mass-loss rates and their frequency of occurrence.
\item The physical mechanism responsible for the blue ward evolution has an important impact on the surface velocities in the yellow and blue supergiant domain: when it is due to a mirror effect (core expansion+envelope contraction), high surface velocities are expected, while when the blue ward evolution is due to strong mass-loss, very low surface velocities are expected.
\end{enumerate}

It would be interesting to explore the impact of a change of the mass-loss rate during the RSG phase at other metallicities. 
Actually, 
we can already say that models computed with the same physics as the present ones but for lower values of Z, would not much be affected by a change
of the mass-loss rate during the RSG phase.The reason is that the present metal poor
models burn most of the helium in the core in the blue region of the HRD \citep[see the discussion in][]{Z002G2013}. Thus
when they enter the RSG phase, they are at the end of their core helium burning stage and thus an increased mass-loss during the very short RSG phase has nearly no effect. However,
\citet{PVI2001} and \citet{Meynet2013} showed that for a different choice of the diffusion coefficients describing the rotational mixing, stars at low metallicity burn most of their helium in the core in the RSG phase.
Enhancing the RSG mass-loss rates in those models would have impacts on the RSG lifetimes and post RSG evolution qualitatively similar to those discussed here. So we see that the discovery of
yellow or blue SN progenitors at metallicities lower or equal than about $Z$=0.006, in the mass range between 15 and 25 M$_\odot$, like the progenitor of SN87A \citep{Arnett1989}, might indirectly provide some hints not only on the physics of mass-loss but also on the physics of rotational mixing!

%========================================================================
\begin{acknowledgements}
The authors thanks Hideyuki Saio for his careful reading of the manuscript, as well as Luc Dessart and Ben Davies for interesting suggestions.
G.M. acknowledges support from the Swiss National Science Foundation (project number 200020-146401) and thanks the hospitality of the Lowell Observatory where part of this work was done.
C.G. acknowledges support from the European Research Council under the European Union Seventh Framework Programme (FP/2007-2013) / ERC Grant Agreement n. 306901.
PM's involvement was supported by the National Science Foundation through AST-1008020.
\end{acknowledgements}
%========================================================================

%We are still far from having a good explanation for the variation with the metallicity of the blue to red supergiant ratio. 
%Many interesting observational facts can be recalled in this context. First the ratio of blue to red supergiant vary a lot depending on the metallicity in clusters with ages between
%NN and NN (Meylan \& Maeder, Eggenberger). In galaxies (REF). 
%A high mass-loss rate during a RSG phase favors a return into bluer regions of the HRD. One can wonder whether there exist some hints that such stars exist.
%discuter effet du mŽlange 
% discuter effet ˆ aigle mŽtallicitŽ

%========================================================================

%\begin{figure*}
%: fig HRDZ014_NH_norot.eps
%\centering
%\includegraphics[width=.98\textwidth]{fig/HRDZ014_NH_norot.eps}
%   \caption{Hertzsprung-Russell diagram for the non rotating models. The colour scale indicates the surface number abundance of nitrogen in a log scale where the abundance of hydrogen is 12. Once the star has become a WNE, the tracks are drawn with black dotted lines.}
%      \label{FigHRDnorot}
%\end{figure*}

%========================================================================

\bibliographystyle{aa}
\bibliography{ENHANCED}
%========================================================================

%========================================================================

\end{document}